\documentclass[sigconf]{acmart}

\usepackage{booktabs} 
\usepackage{amssymb}
\usepackage{hyphenat}
\usepackage{xspace}
\usepackage{subfigure}
\usepackage{multirow}
\usepackage{hhline}
\usepackage{url}

\newcommand{\hide}[1]{}

\newcommand{\xhdr}[1]{\vspace{1.7mm}\noindent{{\bf #1.}}}

\newcommand{\ie}{\textit{i.e.}\xspace}
\newcommand{\eg}{\textit{e.g.}\xspace}
\newcommand{\cf}{\textit{cf.}\xspace}
\newcommand{\etal}{\textit{et al.}\xspace}
\newcommand{\vs}{\textit{vs.}\xspace}
\newcommand{\etc}{\textit{etc.}\xspace}

\newcommand{\Secref}[1]{Sec.~\ref{#1}}

\newcommand{\Tabref}[1]{Table~\ref{#1}}

\newcommand{\Figref}[1]{Fig.~\ref{#1}}

\newcommand{\indep}{{\perp\!\!\!\perp}}

\newcommand{\BA}{BeerAdvocate\xspace}
\newcommand{\RB}{RateBeer\xspace}
\newcommand{\fr}{first\hyp rating\xspace}
\newcommand{\ptg}{paired\hyp treatment group\xspace}

\begin{document}

\copyrightyear{2018}
\acmYear{2018}
\setcopyright{iw3c2w3}
\acmConference[WWW'18]{The 2018 Web Conference}{April 23--27, 2018}{Lyon, France}
\acmBooktitle{WWW 2018: The 2018 Web Conference, April 23--27, 2018, Lyon, France}
\acmPrice{}
\acmDOI{10.1145/3178876.3186160}
\acmISBN{978-1-4503-5639-8}

\title{When Sheep Shop: Measuring Herding Effects in Product Ratings with Natural Experiments}

\author{Gael Lederrey}
\affiliation{%
  \institution{EPFL}
}
\email{gael.lederrey@epfl.ch}

\author{Robert West}
\orcid{0000-0002-3984-1232}
\affiliation{%
  \institution{EPFL}
}
\email{robert.west@epfl.ch}

\begin{abstract}
As online shopping becomes ever more prevalent, customers rely increasingly on product rating websites for making purchase decisions.
The reliability of online ratings, however, is potentially compromised by the so-called \textit{herding effect:}
when rating a product, customers may be biased to follow other customers' previous ratings of the same product.
This is problematic because it skews long-term customer perception through haphazard early ratings.
The study of herding poses methodological challenges.
In particular, observational studies are impeded by the lack of counterfactuals: simply correlating early with subsequent ratings is insufficient because we cannot know what the subsequent ratings would have looked like had the first ratings been different.
The methodology introduced here exploits a setting that comes close to an experiment, although it is purely observational---a \textit{natural experiment}.
Our key methodological device consists in studying the same product on two separate rating sites, focusing on products that received a high first rating on one site, and a low first rating on the other.
This largely controls for confounds such as a product's inherent quality, advertising, and producer identity, and lets us isolate the effect of the first rating on subsequent ratings.
In a case study, we focus on beers as products and jointly study two beer rating sites, but our method applies to any pair of sites across which products can be matched.
We find clear evidence of herding in beer ratings.
For instance, if a beer receives a very high first rating, its second rating is on average half a standard deviation higher, compared to a situation where the identical beer receives a very low first rating.
Moreover, herding effects tend to last a long time and are noticeable even after 20 or more ratings.
Our results have important implications for the design of better rating systems.
\end{abstract}

\maketitle

\section{Introduction}
\label{sec:intro}

With every purchase but one click away, online shopping is extremely convenient and is accounting for an ever larger share of the retail market.
The downside of online shopping is that it provides a less direct experience than going to an off\/line, brick\hyp and\hyp mortar store, where customers can taste, smell, touch, and feel a product before deciding whether to buy it.
Online, we must rely on ratings provided by previous customers instead.

Online rating systems, however, suffer from the known problem of social influence, also termed \textbf{herding,} which expresses the fact that raters tend to be biased by the opinions of previous raters \cite{chevalier2006effect,glenski2017rating,hu2006can,lee2015follow,muchnik2013social,wang2014amazon} and which can make online rating systems fickle and sensitive to small variations in early ratings: if the first few reviews of a product happen to swing a certain way (or are purposefully engineered that way in an act of review spamming \cite{jindal2007analyzing}), this can unduly skew subsequent reviews.

\begin{figure}[t]
    \centering
    \subfigure[Ratings]{
        \includegraphics[width=0.45\columnwidth]{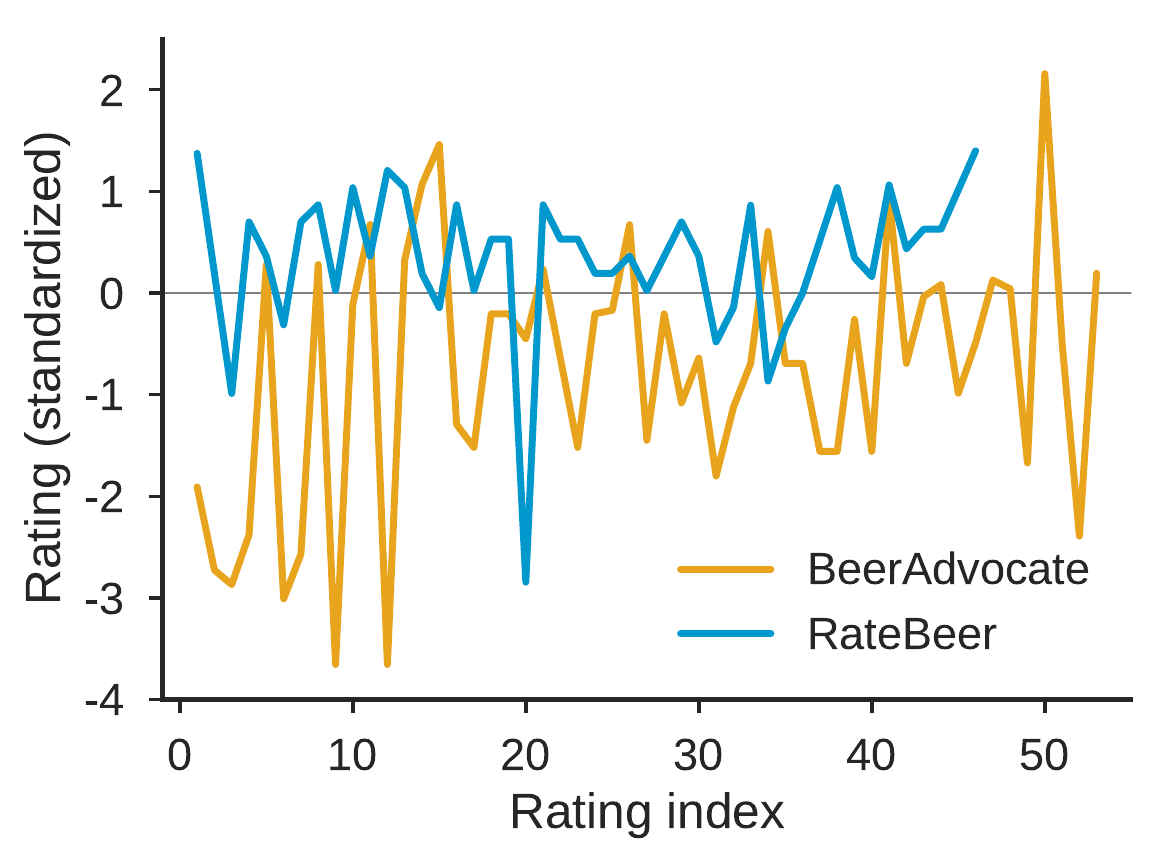}
        \label{fig:example time series}
    }
    \subfigure[Cumulative average ratings]{
        \includegraphics[width=0.45\columnwidth]{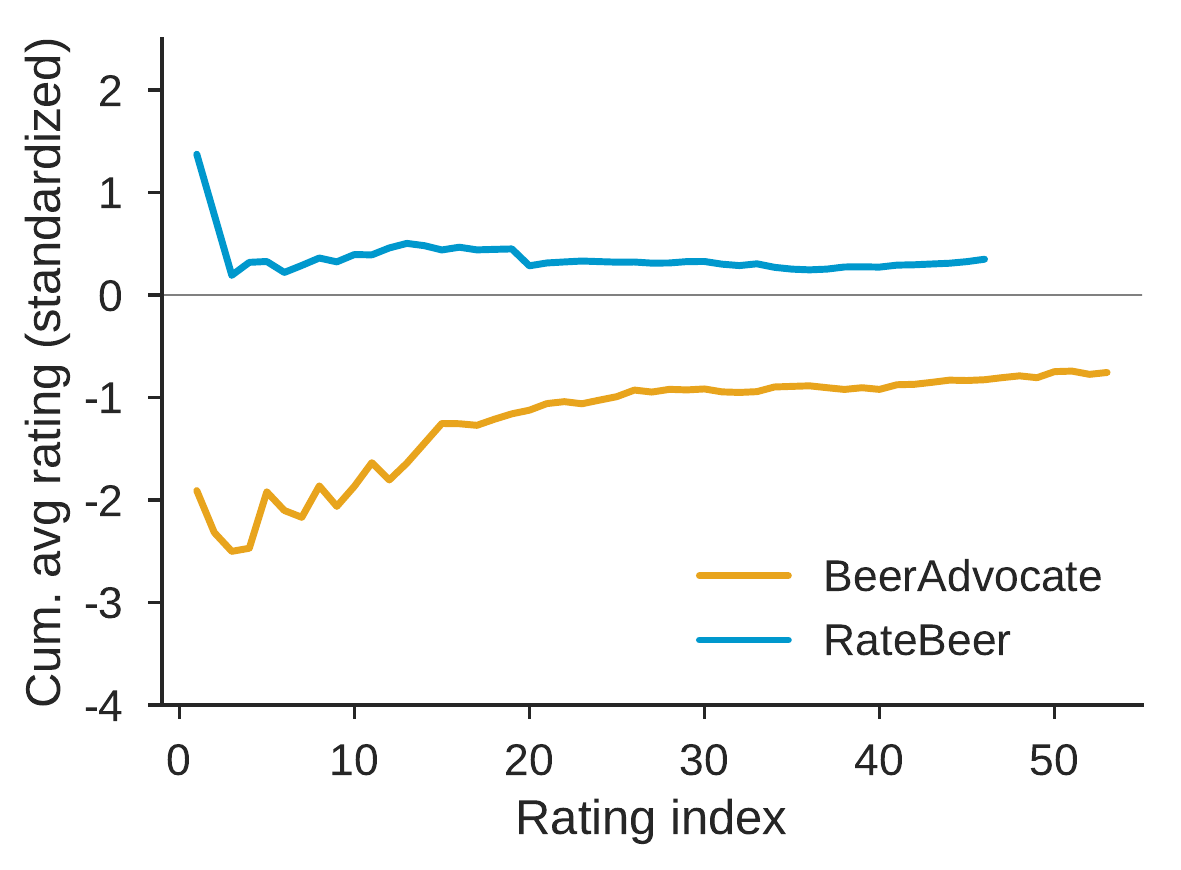}
        \label{fig:example running avg}
    }
    \vspace{-5mm}
    \caption{
    \textit{Lost Rhino Ice Breaker IPA,} an example of a beer with low (high) ratings on \BA (\RB)
    (scores standardized to be comparable across sites; \cf \Secref{sec:basic analysis}).
    }
    \label{fig:example}
    \vspace{-3mm}
 \end{figure}


This can have severe implications for both customers, who may end up with unsatisfying products, and producers, whose high\hyp quality products may end up being bought less than they deserve.
Herding is therefore a behavior of great interest both sociologically as well as economically.


Studying the herding effect is difficult, however.
Although \textbf{randomized experiments} have been successfully deployed to quantify social influence in rating behavior \cite{glenski2017rating,muchnik2013social}, experiments can be risky from a business and ethical perspective, for either they wilfully subject products to random treatments with potentially harmful effects, or they are restricted to small laboratory settings \cite{roider2016reputational}, which reduces the generality of findings.
\textbf{Observational studies} of herding \cite{chevalier2006effect,duan2008online,lee2015follow}, on the other hand, are less delicate in this regard than randomized experiments, but they are more delicate methodologically, as they require more care during the analysis.


To illustrate this point, consider a \textbf{na\"ive observational study of herding,}
which simply measures if products that receive high early ratings also tend to receive high later ratings, and if at the same time products that receive low early ratings tend to receive low later ratings.
The problem, obviously, with this hypothetical study is that both early and later ratings might be caused by a hidden correlate---the inherent quality of the product---rather than one by the other.
To credibly claim a causal relation, we would need to show that, \textit{for a fixed product,} later ratings follow early ratings regardless of whether the first ratings were high or low.
If, on the contrary, inherent quality were the real cause, then a good product would get good later ratings even if, for some reason, the first ratings were low.
In other words, the na\"ive observational study suffers from the lack of counterfactuals: we cannot tell whether there has been herding in the ratings for a product if we do not know what would have happened if the first ratings had been different.

    
We address this challenge by studying not a single product rating website, but two of them in parallel.
In particular, we will observe how the same product is rated independently on the two sites;
if the product happens to receive a vastly different \fr on site 1, compared to site 2, we can measure how the \fr affects subsequent ones.

To explain the basic idea behind our method, we start with an example.
For the sake of concreteness, and to set the stage for our case study, consider the case of beers as products, rated on the two major beer\hyp rating websites, \BA and \RB.
Consider a beer $B$ that has become available just recently. The \RB user to first rate beer $B$ on \RB happens to love it, so she gives it a very high score. The \BA user to first rate beer $B$ on \BA, on the contrary, happens to hate it, so he gives it a very low score.
Of course, the inherent quality of beer $B$ was exactly the same for both users, and each user was the first to rate beer $B$ on the respective site, so they were not influenced by previous opinions.
It is only due to chance that the first \RB rating was high, and the first \BA rating low, instead of
\textit{vice versa.}

More generally, whenever the same beer gets a high \fr on one site, and a low \fr on the other site, it seems likely that it is haphazard whether the high \fr happens on \RB or on \BA.
Whether this is indeed true needs to be established, but if
it can be established,
this means that nature has created a situation akin to an experiment for us: she has flipped a coin to decide on which of the two sites beer $B$ was to be exposed to the ``experimental'' condition \textit{high \fr,} and on which to the ``experimental'' condition \textit{low \fr}---a so-called \textbf{natural experiment}.
We may then analyze which effect the two conditions have on later ratings of beer $B$. In the presence of herding, later ratings on the site with the high \fr should on average be higher than on the site with the low \fr. Conversely, in the absence of herding, later ratings should on average be similar regardless of the \fr.

\Figref{fig:example} shows that beers such as the anonymous $B$ really exist, in this case an India pale ale named \textit{Lost Rhino Ice Breaker}.
The time series of this beer's ratings is plotted in \Figref{fig:example time series}, with the cumulative average displayed in \Figref{fig:example running avg}.
The curves show that \textit{Lost Rhino Ice Breaker} received a much lower \fr on \BA than on \RB, and that it never managed to recover from its dismal start on \BA, while it thrived on \RB---a difference that may be due to herding.%
\footnote{The irony of illustrating herding with a lost rhino is incidental.}
This example is but anecdotal, of course, and the goal of this paper is to move from such examples to reliable causal statements.
Our main contributions are twofold:
First, we introduce an observational methodology for quantifying how consistently early ratings influence later ones via the natural experiment sketched above (\Secref{sec:methodology}).
Second, we apply our method to a real dataset of product ratings (\Secref{sec:data}), after carefully ruling out confounds and thus confirming that indeed we have identified a natural experiment (\Secref{sec:matching}).
Our results provide strong evidence of substantial herding effects (\Secref{sec:results}): the second rating for a product is on average half a standard deviation higher when a beer received a very high \fr, compared to when it received a very low \fr.
Furthermore, herding effects are tenacious and can be noticed even after 20 or more reviews.
We conclude the paper by discussing implications of our findings and prior as well as future work (\Secref{sec:discussion}).

\section{Methodology: natural experiment}
\label{sec:methodology}

\begin{figure}[t]
    \centering
    \footnotesize
\begin{tabular}{cc|cc}
    \hspace{-3mm}
    \subfigure{
        \includegraphics[width=0.22\columnwidth]{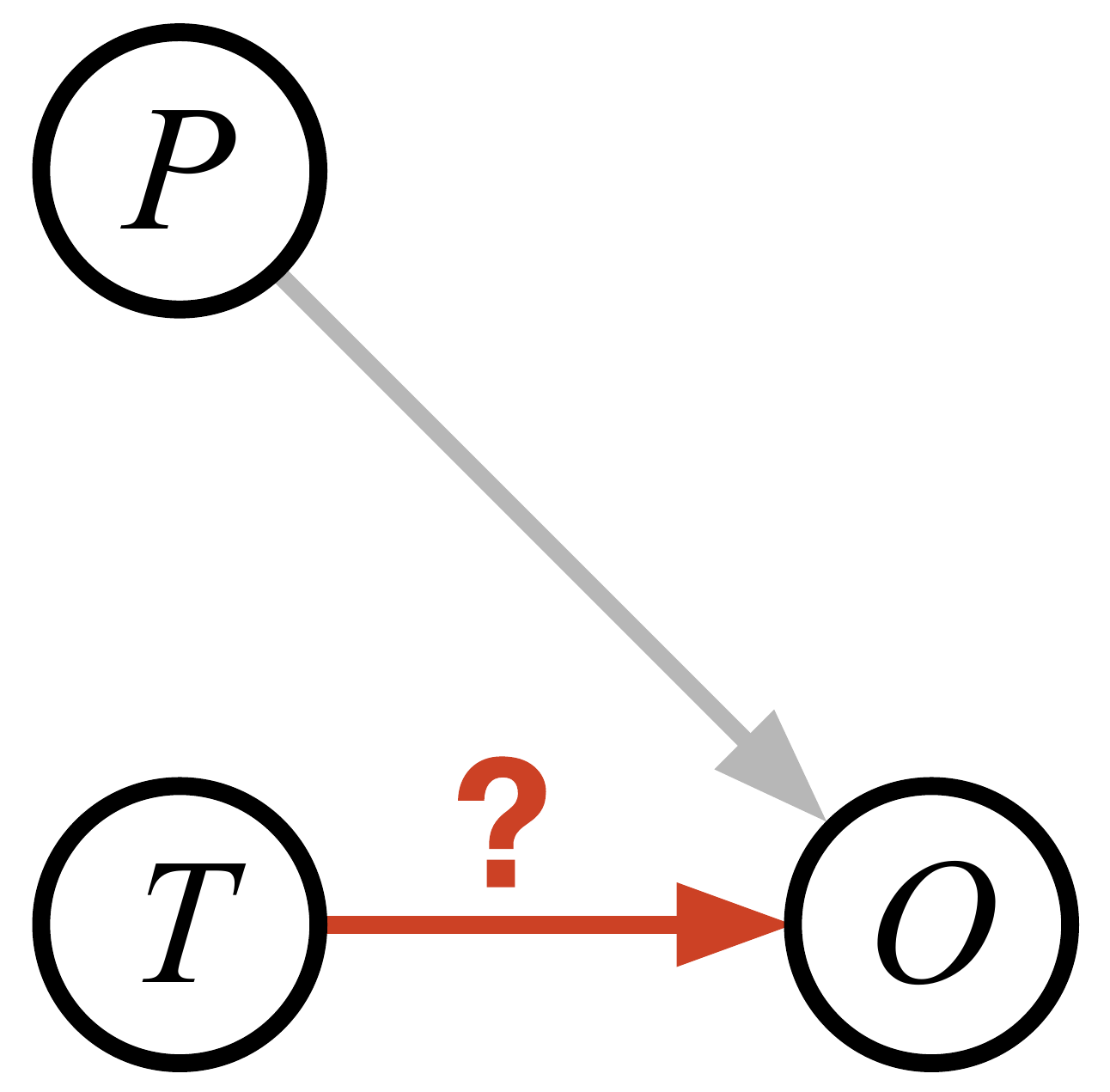}
        \label{fig:bayesian_networks_exp}
    }
&
    \hspace{-3mm}
    \subfigure{
        \includegraphics[width=0.22\columnwidth]{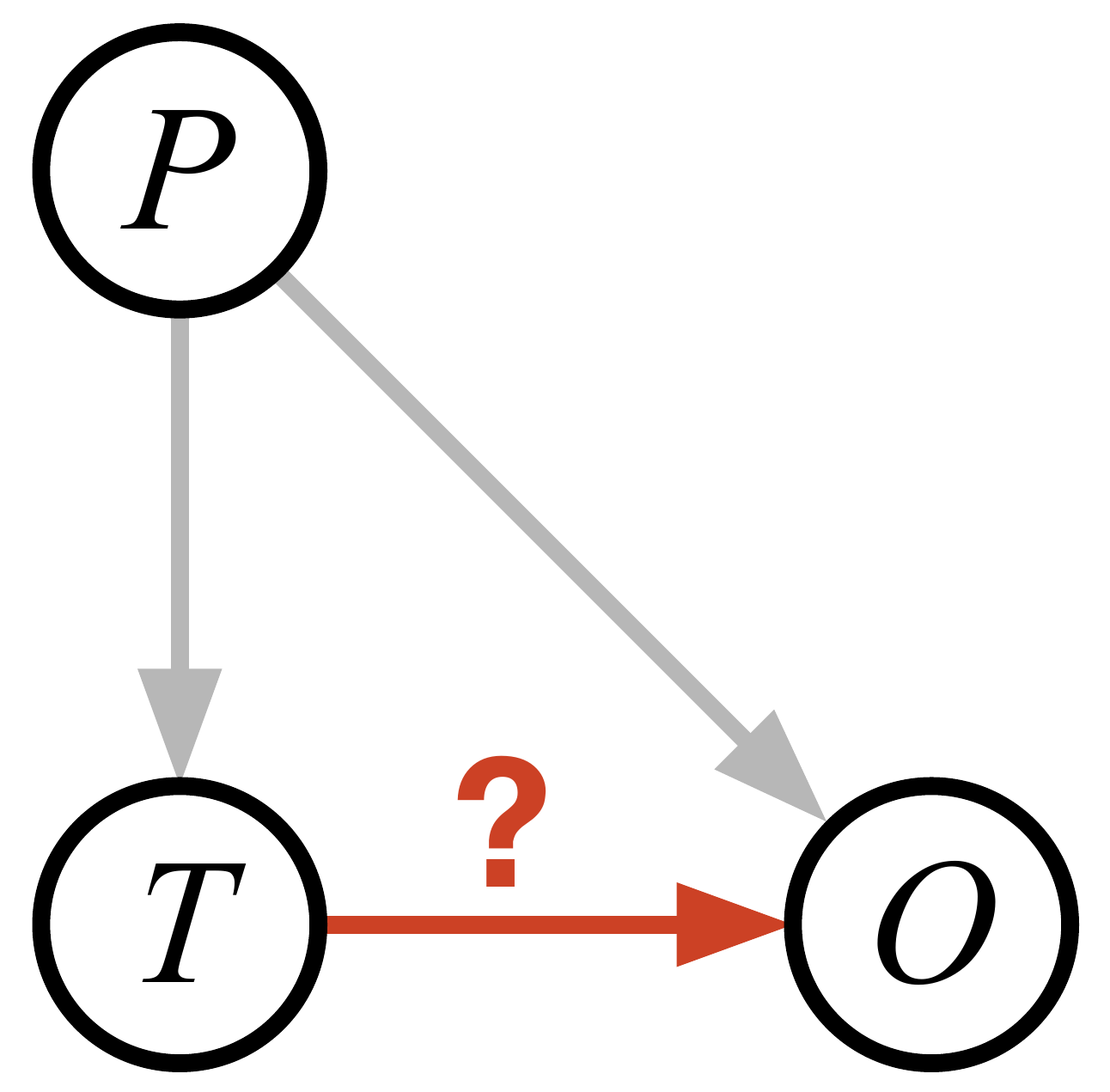}
        \label{fig:bayesian_networks_naive}
    }
&
    \subfigure{
        \includegraphics[width=0.22\columnwidth]{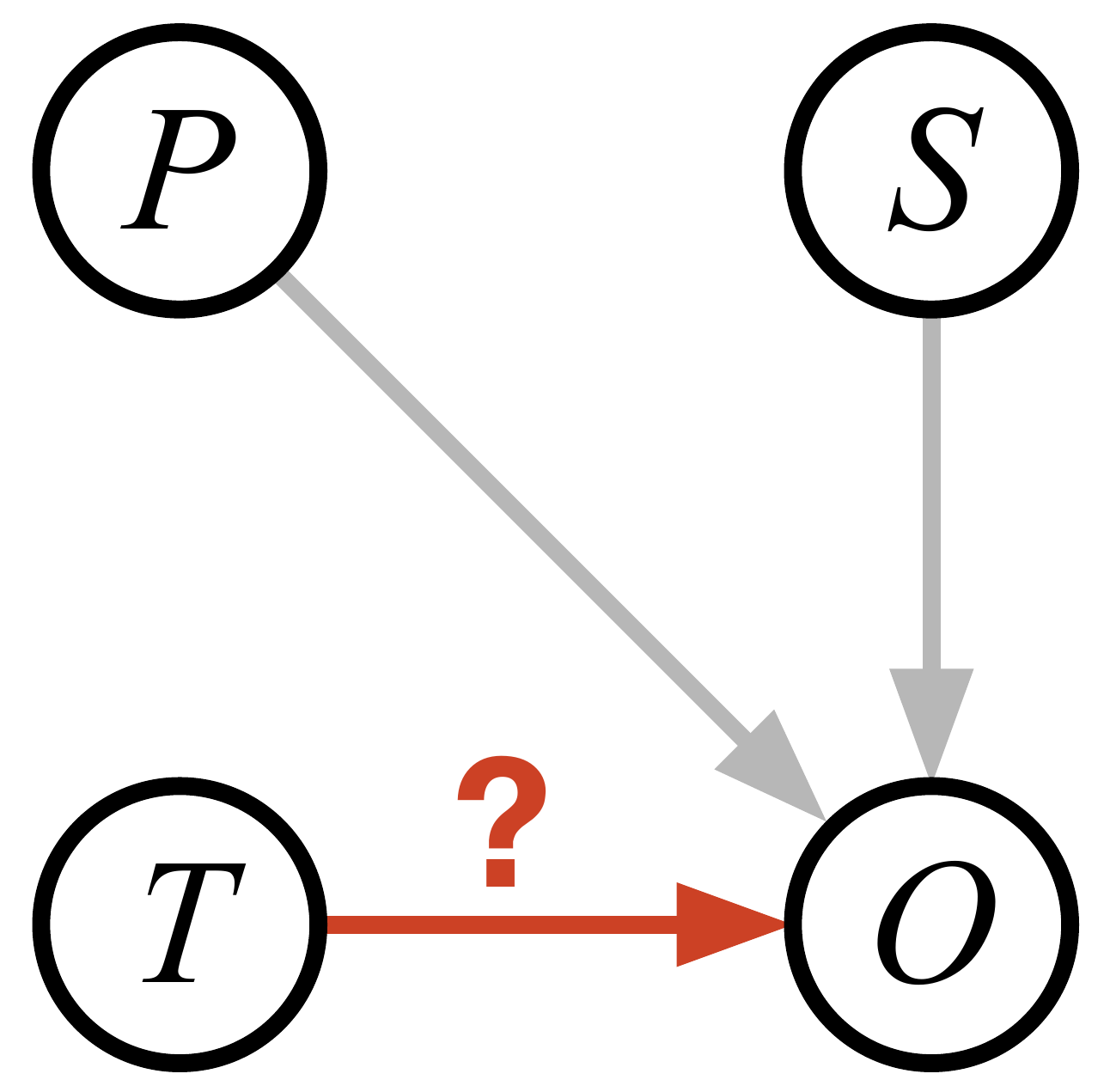}
        \label{fig:bayesian_networks_good_natexp}
    }
&
    \hspace{-3mm}
    \subfigure{
        \includegraphics[width=0.22\columnwidth]{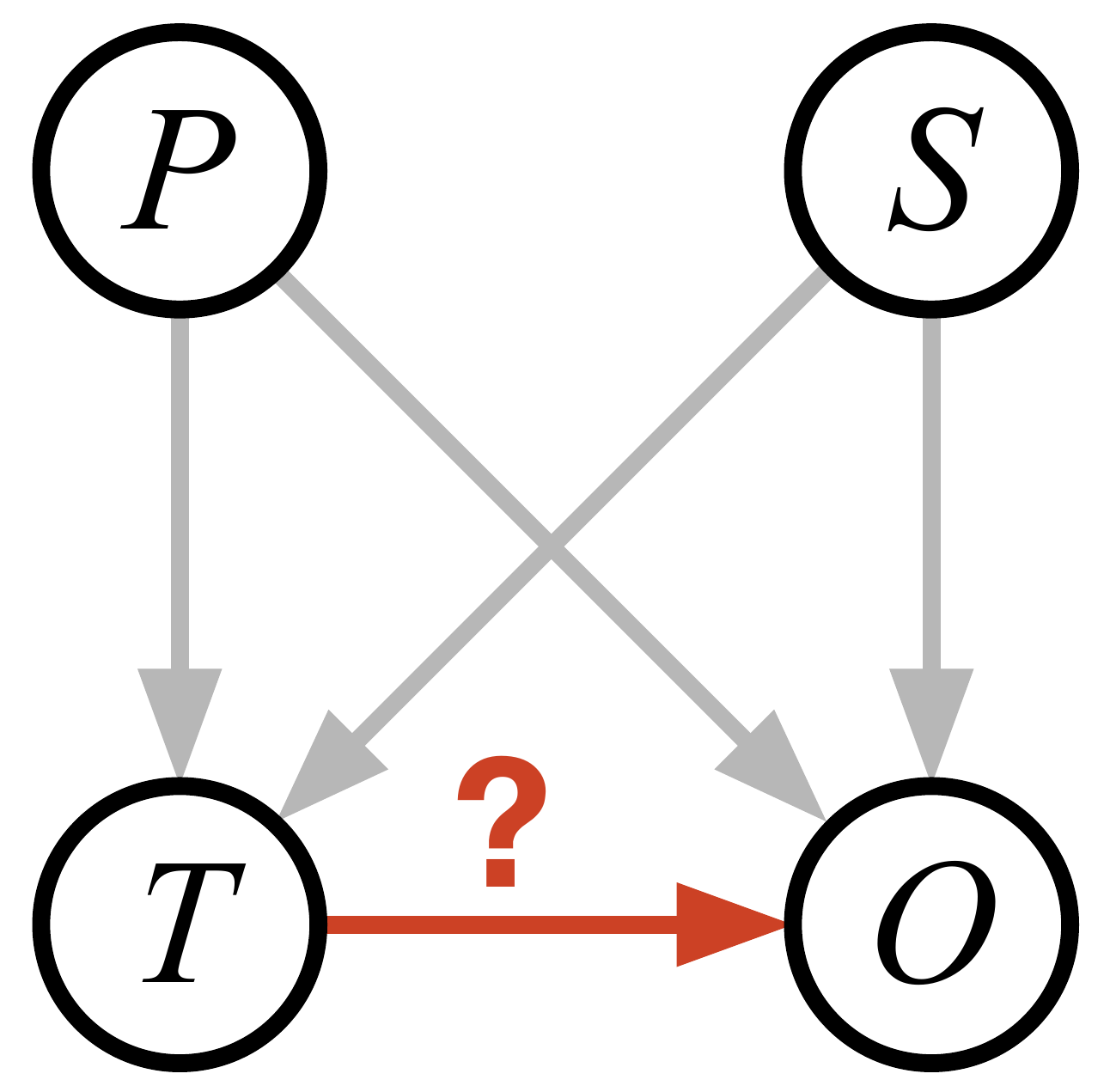}
        \label{fig:bayesian_networks_bad_natexp}
    }
\\
(a)~Randomized 
& \hspace{-3mm}
(b)~Na\"ive observ-
&
(c)~Good natural
& \hspace{-3mm}
(d)~Bad ``natural 
\\
experiment 
& \hspace{-3mm}
ational study 
&
experiment
& \hspace{-3mm}
experiment''
\end{tabular}
    \vspace{-3mm}
    \caption{
    Variable dependencies for different kinds of study (\Secref{sec:methodology}).
    \textbf{\textit{T}}: treatment (\fr);
    \textbf{\textit{O}}: outcome (subsequent ratings);
    \textbf{\textit{P}}: rated product;
    \textbf{\textit{S}}: rating site.
    Only (a) and (c) allow us to draw conclusions about a causal link between \textbf{\textit{T}} and \textbf{\textit{O}}.
    }
    \label{fig:bayesian_networks}
    \vspace{-5mm}
 \end{figure}

To illustrate the difference between a proper randomized experiment and the na\"ive observational study delineated in the introduction, we represent both scenarios as Bayesian networks in \Figref{fig:bayesian_networks}(a--b).
The diagrams contain three random variables:
product ($P$), treatment ($T$), and outcome ($O$).
The treatment $T$ captures whether the \fr for product $P$ is high or low;
the outcome $O$ captures whether subsequent ratings for $P$ are high or low.

In a randomized experiment (\Figref{fig:bayesian_networks_exp}), treatment assignment is decided by a coin flip, independent of any product properties, so different treatment groups are indistinguishable with respect to product properties and are therefore directly comparable.
Hence, if we observe significantly different outcomes for different treatments, this difference is likely to be caused by the treatment.

In a na\"ive observational study (\Figref{fig:bayesian_networks_naive}), on the contrary, treatment assignment may depend on product properties, so the latter may influence both treatment and outcome.
In this case, a correlation between treatment and outcome may not be causal, but both may instead be caused separately by a \textit{confound,} a latent property of the product.
As mentioned, an obvious confound could be the inherent quality of the product, as good products tend to be rated highly both by the first reviewer (treatment) and by subsequent reviewers (outcome), even in the absence of herding.

Our methodology circumvents the problem of product\hyp induced confounds by studying not one single rating website (as would be done in a na\"ive observational study), but rather two separate rating websites with overlapping sets of rated products.
This still constitutes an observational study, but one that naturally controls for confounds and thus comes close to a randomized experiment in spirit---a situation commonly known as a \textit{natural experiment.}

Let us call the two sites $S_1$ and $S_2$, and consider a product $P$
rated on both sites.
If $P$ received a high \fr (\ie, treatment) on $S_1$ and a low \fr on $S_2$ (or \textit{vice versa}), then---under conditions to be discussed below (\Secref{sec:assumptions})---this situation may be seen as emulating two ``parallel universes'': we can observe what happens to the \textit{same} product under each possible treatment.
In other words, we now have counterfactuals, the lack of which is the major shortcoming of the na\"ive, single\hyp site observational study.

\vspace{-1mm}

\subsection{Step-by-step description}
\label{sec:steps}

We now describe our methodology in detail. It encompasses 5 steps.

\textbf{Step~1:} \textbf{Match products} across the two rating websites.
The fundamental device exploited by our natural experiment consists in tracking the same product on two separate websites, so we work only with products that we can identify on both sites and discard all others.
\Secref{sec:matching_algorithm} describes the matching algorithm we used on a pair of beer rating websites; it is based on string similarities between product names, and we believe it is general enough to be adapted to other datasets as well.

\textbf{Step~2:} \textbf{Define the paired\hyp treatment groups.}
On each site, label each product as ``high'' (H), ``medium'' (M), or ``low'' (L), depending on the \fr (treatment) it received on the site.
A \fr is defined as H if it is in the top $p$ percent of the \fr{}s on the respective site, as L if it is in the bottom $p$ percent, and as M otherwise.
(In our specific case study, we use $p=15$.)
After this step, each product falls into one of the nine \textit{paired\hyp treatment groups} defined by the cross product
$\{ T_1 T_2 : T_1,T_2 \in \{\text{H}, \text{M}, \text{L}\}\}$, where the two letters capture the treatments the product received on sites $S_1$ and $S_2$ via the respective \fr{}s;
\eg, HH means that the product received high \fr{}s on both sites, HL that it received a high \fr on $S_1$ and a low one on $S_2$, LH that it received a low \fr on $S_1$ and a high \fr on $S_2$, \etc

\textbf{Step~3:} \textbf{Balance the paired\hyp treatment groups,} making sure that, for each $(T_1,T_2) \in \{\text{H}, \text{M}, \text{L}\} \times \{\text{H}, \text{M}, \text{L}\}$, we have the same number of products in the paired\hyp treatment group $T_1 T_2$ as in the group $T_2 T_1 $.
We may achieve this simply by randomly subsampling from the larger of the two groups.
This step ensures that, for each site $S$ and each \ptg $T_1 T_2$ (with $T_1 \neq T_2$), the probabilities of the two treatments $T_1$ and $T_2$ are 50\% each.

\textbf{Step~4:} \textbf{Aggregate paired\hyp treatment groups} containing the same set $\{T_1,T_2\}$ of treatments.
This reduces the number of paired\hyp treatment groups to six: HH, HM, HL, MM, ML, LL;
\eg, after aggregation, the group HL contains both products with H on $S_1$ and L on $S_2$ and products with L on $S_1$ and H on $S_2$.
This is done to have more data points per group, but we advise to also perform a separate analysis on the non\hyp aggregated data as a sanity check.

\textbf{Step~5:} \textbf{Compare the outcomes for different treatments} within the same \ptg.
We consider the groups HL, HM, and ML, where the same product received different \fr{}s on the two sites.
By comparing subsequent ratings across the two sites, we can estimate the treatment effect in isolation from any product\hyp related confounds (such as inherent quality), which are controlled for by fixing the product.
In particular, for a given product $P$ and a given rating index $i$, we compare $P$'s $i$-th rating on the site where it received the higher \fr with its $i$-th rating on the site where it received the lower \fr.
If the difference is positive, this supports the hypothesis of a causal link between treatment (\fr) and outcome ($i$-th rating), \ie, herding.
Tracking the difference as a function of the rating index $i$ also lets us study if, and how fast, herding attenuates with time.

The most interesting \ptg is HL, as it corresponds to the starkest difference in treatments.
Since, however, it will generally occur less frequently in practice than the less extreme groups (HM and ML), we recommend to also study the latter.
Finally, as a sanity check, it is also advisable to include the symmetric groups HH, MM, and LL in the analysis.

\subsection{Assumptions}
\label{sec:assumptions}

The above methodology allows us to estimate the causal effect of the \fr (treatment) on subsequent ratings (outcome) if the two following assumptions hold.

The first and most crucial assumption, and in fact the defining property of a natural experiment, is that \textbf{treatment assignment is haphazard:} whether a given product $P$ with different \fr{}s receives its higher \fr on $S_1$ and its lower \fr on $S_2$ or \textit{vice versa} must not depend on any properties of $P$, $S_1$, and $S_2$.
In other words, the treatment assignment $T$ must be independent of the product $P$ and of the site $S$ (\Figref{fig:bayesian_networks_good_natexp}).
If this is the case, and if variations in $T$ are correlated with variations in $O$, then the link between $T$ and $O$ is likely to be causal.
Otherwise (\Figref{fig:bayesian_networks_bad_natexp}), properties of the product or of the site, or a combination of the two, could explain both the treatment assignment and the outcome---we might have mere correlation without causation, which would defeat the very purpose of considering a matched, rather than a na\"ive, single-site observational study.

Note that, by construction, treatment assignment is independent of the product alone: each product is included twice in the matched dataset, once per site, on one of them with a higher, and on one with a lower, \fr, such that we have a 50/50 distribution over the two possible treatments for a given product.
Similarly, again by construction, treatment assignment is independent of the site alone: after balancing \ptg{}s (step~3 in \Secref{sec:steps}), each site has as many products with a higher as with a lower \fr, resulting in a 50/50 distribution over the two possible treatments in each \ptg for a given site.

This does not imply, though, that treatment assignment is independent of the combination of product and site;%
\footnote{While $X \indep (Y,Z)$ (``$X$ is independent of $(Y,Z)$'') implies $X \indep Y$ and $X \indep Z$, the statement does not hold in the opposite direction.}
\eg, users on site $S_1$ might like a certain kind of product more than users on site $S_2$, which could result in an increased probability of a higher \fr{} (treatment) 
for that kind of product on $S_1$, compared to $S_2$.

In our setup, showing that treatment is indeed independent of the combination of product and site establishes the
\textit{internal validity} of the study.
How to show that a specific study is internally valid depends on the datasets being used. (See \Secref{sec:validity} for how we proceed in the case of beer ratings.)
Although we cannot speak of a natural experiment if this independence does not immediately hold in the matched dataset, one might still achieve it by explicitly balancing the dataset, \eg, via propensity\hyp score matching \cite{rosenbaum2002observational}.

The second assumption is that the \textbf{matched dataset accurately reflects the full dataset.}
In general, not all products are present on both rating sites, so matching will select a subset of all products.
If the matched sample is biased, \ie, systematically different from the full dataset before matching, this might preclude us from generalizing our findings from the natural experiment to the set of all products rated on the two sites.
For instance, it is conceivable that particularly good products are more likely to be present on both sites, which would make our findings specific to good, rather than average, products.
By showing that the matched sample is unbiased, we establish the so-called \textit{external validity} of the study. (See \Secref{sec:validity} for how we proceed in the case of beers as products.)


\section{Data: two beer rating websites}
\label{sec:data}
We apply our methodology to the specific scenario of beer ratings.
This setting is well suited for several reasons:
the market is dominated by two large websites dedicated to the rating and reviewing of beers---\BA and \RB---, each with a long history reaching back nearly 20 years, with very similar site designs, and with a large overlap of rated products.

An older version of the data was made available by McAuley \textit{et~al.}\ \cite{leskovec2016snap, McAuley2012LearningAA}, but as that version was produced in 2012, we re\hyp crawled it; the data now extends from 2001 to August 2017.%
\footnote{
Data available upon request.
Code: \href{https://github.com/epfl-dlab/when_sheep_shop}{\url{https://github.com/epfl-dlab/when_sheep_shop}}.
}

Although we focus on one case study, our method applies equally to other pairs of rating websites, as long as the intersection of the sets of rated products is large (\cf our discussion in \Secref{sec:discussion}).

\subsection{Description of beer rating websites}
\label{sec:site description}

\BA and \RB are the two largest online beer\hyp related websites.
Although they provide a general space for beer aficionados, with articles, discussion forums, and trading platforms, their main purpose is to collect and curate beer ratings provided by users.
On both sites, beers are rated with respect to five aspects (look, smell\slash aroma, taste, feel\slash palate, overall), which are then combined via a weighted sum into a rating score between 1 and 5.
The sites are also similar with respect to layout and visual appearance.
They both prominently show the most recent ratings as well as the current cumulative average on the page of each beer (\RB also shows the rank of the beer among all beers), so we can assume that users about to rate a beer become aware of this information.

\subsection{Basic analysis of rating datasets}
\label{sec:basic analysis}

Here we discuss some properties of the two beer rating datasets that are relevant for our study of herding.

We start by summarizing the size of the datasets in \Tabref{tbl:dataset size}, which shows that each site contains ratings for hundreds of thousands of beers from tens of thousands of breweries, rated by tens of thousands of users, totaling millions of ratings.

\begin{table}[t]
  \caption{Dataset size.}
  \vspace{-3mm}
  \label{tbl:dataset size}
  \begin{tabular}{l|rr}
    & BeerAdvocate & RateBeer \\ \hline
    Breweries & 16,758 & 24,189 \\ \hline
    Beers & 280,823 & 442,081 \\
    Beers ($\geq$ 5 ratings) & 96,156 & 166,043 \\
    Beers ($\geq$ 10 ratings) & 61,193 & 104,062 \\
    Beers ($\geq$ 20 ratings) & 38,533 & 60,451 \\ \hline
    Users & 153,704 & 70,174 \\
    Users ($\geq$ 10 ratings) & 48,595 & 17,744 \\
    Users ($\geq$ 100 ratings) & 14,488 & 6,419 \\ \hline
    Ratings & 8,393,032 & 7,122,074 \\
  \end{tabular}
  \vspace{-5mm}
\end{table}


The two sites attract rather different user populations.
In particular, \BA is mostly frequented by users from the U.S.\ (74\% of users), followed by Canada (2\%), with less than 1\% of users from any other single country.
\RB's user base, though also predominantly from the U.S.\ (38\%), is more balanced, with 5\% of users from Canada, 4\% from England, 2\% Poland, 2\% from Australia, \etc

\BA's more U.S.\hyp centric user base is also reflected in the breweries whose beers are rated on the sites:
44\% of all breweries represented on \BA are from the U.S., while the fraction is only 29\% on \RB.
Other countries have similar percentages across sites (\cf ``Unmatched'' in \Tabref{tbl:brewery locations}).
These differences imply that selecting a matched sample of beers rated on both websites cannot possibly reflect the overall distribution of products on both sites equally well, an issue we address in \Secref{sec:validity}.

\begin{figure*}[t]
    \vspace{-3mm}
    \centering
    \subfigure[Histograms of ratings]{
        \includegraphics[width=0.31\textwidth]{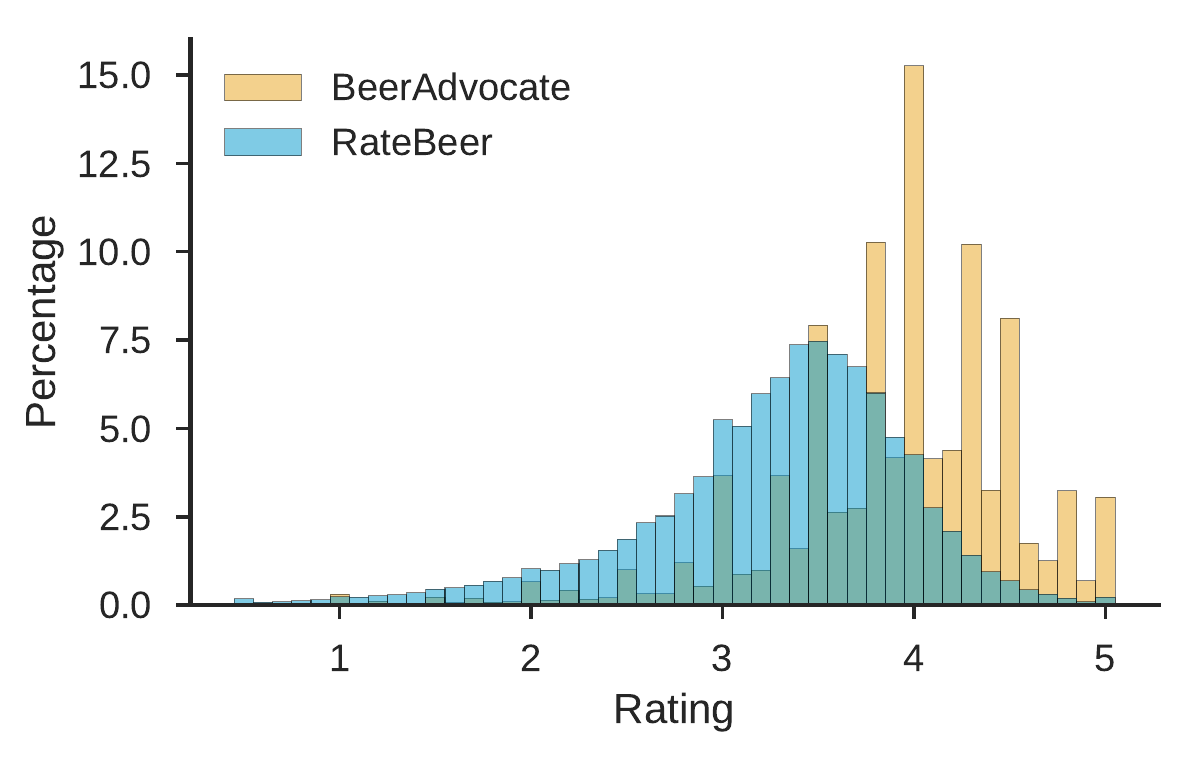}
        \vspace{-2mm}
        \label{fig:raw rating hist}
    }
    \subfigure[Mean of ratings per year]{
        \includegraphics[width=0.31\textwidth]{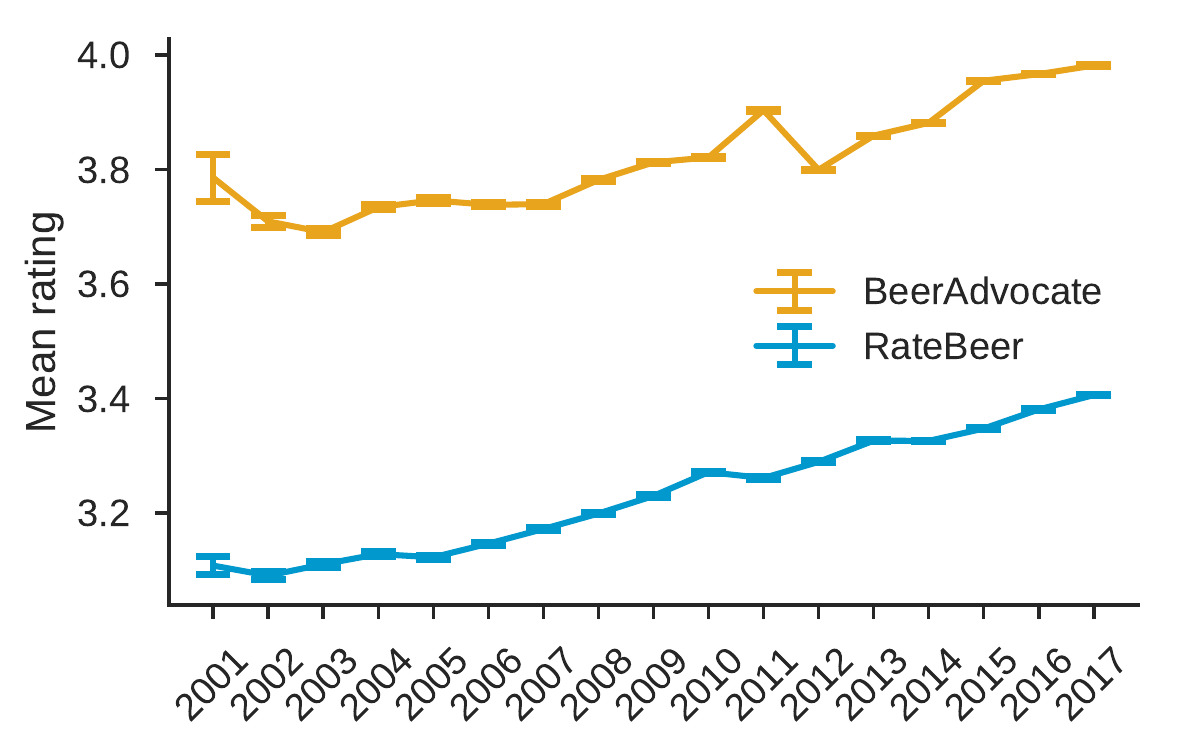}
        \label{fig:average per year}
    }
    \subfigure[Standard deviation of ratings per year]{
        \includegraphics[width=0.31\textwidth]{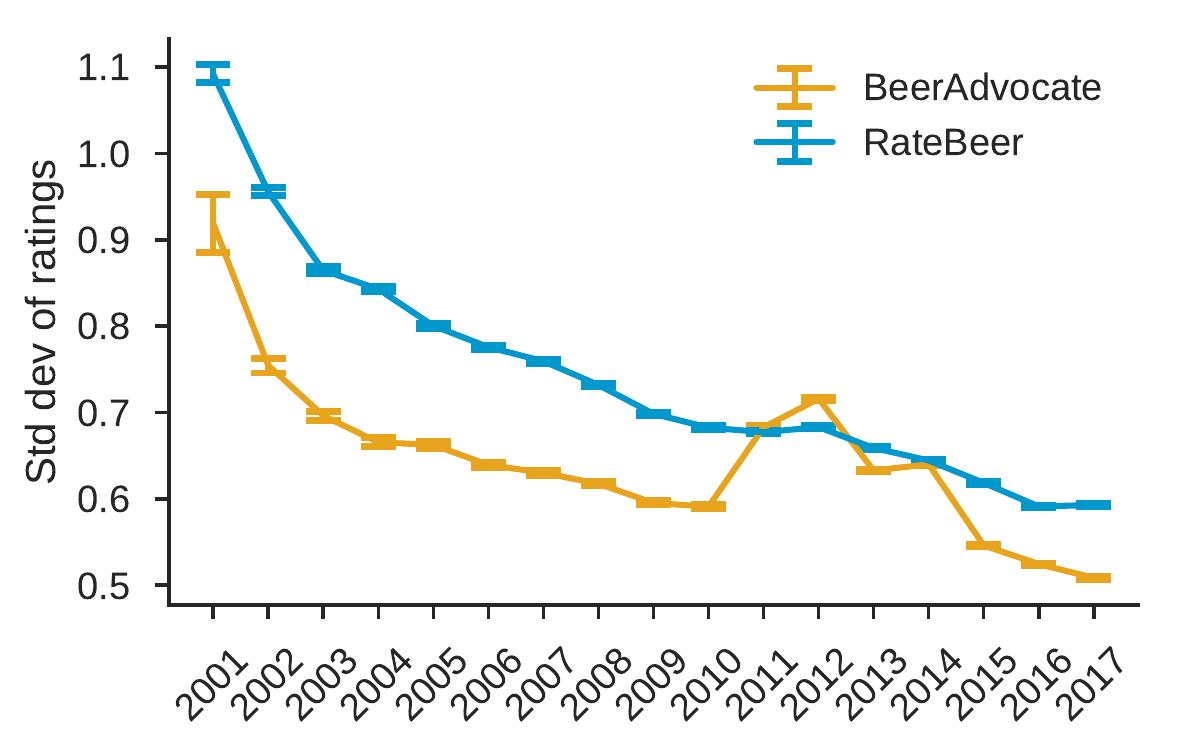}
        \label{fig:std per year}
    }
    \vspace{-4mm}
    \caption{
    Statistics of the two beer\hyp rating datasets. Error bars in (b--c) capture 95\% confidence intervals.
    }
    \label{fig:mean and std per year}
    \vspace{-2mm}
 \end{figure*}

\Figref{fig:raw rating hist} plots the histograms of ratings for both websites.
We clearly see that users on \BA tend to give higher ratings than users on \RB.%
\footnote{We found that the distinct spikes in \BA's histogram are caused by reviews that gave the same score to all five aspects (\cf \Secref{sec:site description}) and that tend to be very short or even empty, which seems to indicate that these reviews were entered in a hurry.}
As a side note, we also point out that the rating distributions of \Figref{fig:raw rating hist} differ vastly from the bimodal distributions with mostly extremely high or extremely low values that have frequently been observed on other rating websites \cite{hu2006can} and that have been attributed to a ``brag-and-moan'' effect.
On beer rating websites, ratings seem to be less affected by selection bias due to disappointment or positive surprise.




The histograms of \Figref{fig:raw rating hist} pool all ratings from 2001 through August 2017.
Next, we group ratings by year and plot the annual mean (\Figref{fig:average per year}) and standard deviation (\Figref{fig:std per year}).
We observe that neither quantity stays constant over time:
the mean increases, while the standard deviation decreases, from year to year.
Assuming that the inherent quality of beers being rated stays roughly constant, the rising mean may be interpreted as score inflation, while the sinking standard deviation could indicate a consolidating consensus about what should constitute the score of an average beer.

This implies that, in order to compare ratings across sites and time periods, we must account for biases stemming from site conventions (shifted rating histograms) and from a temporal drift in these conventions (rising means and sinking standard deviations).
Instead of raw ratings, we therefore consider \textit{standardized ratings} (also known as $z$-scores):
for each site and each year, we compute the mean and standard deviation over all ratings.
We then subtract the mean of year $t$ from all ratings submitted in year $t$ and divide them by the standard deviation of year $t$, such that each year's set of ratings has mean 0 and standard deviation 1.

\section{Matching products across websites}
\label{sec:matching}
Our method hinges on a set of products rated on two separate websites.
An alignment of products across sites (\eg, via consistent unique identifiers) is typically not given explicitly; rather, one usually needs to perform the matching oneself heuristically.
In this section, we describe our algorithm for achieving a high\hyp quality alignment between beers from \BA and \RB (\Secref{sec:matching_algorithm}), report basic statistics of the matched sample (\Secref{sec:matched_dataset}), and discuss its external and internal validity (\Secref{sec:validity}).

\subsection{Matching algorithm}
\label{sec:matching_algorithm}

When matching products across websites, we should favor precision over recall: not matching all potentially matchable products simply decreases the sample size and is therefore acceptable (as long as it does not bias the dataset, \cf \Secref{sec:validity}), whereas matching non\hyp identical products to each other introduces noise into the data.
With this in mind, we designed a matching algorithm geared toward precision, potentially at the expense of recall.

We proceed in two phases.
First, we align breweries across sites, then we align beers only within breweries; \ie, we only consider pairs of beers as potential matches if their respective breweries were matched to each other before.
The same basic procedure is used both for matching breweries and for matching beers, in each case operating on name strings.

The algorithm starts by representing names as TF-IDF vectors and computes all pairwise cosine similarities. Inverse document frequency (IDF) weighting serves to down\hyp weight common terms such as ``brewery'', ``company'', ``beer'', ``ale'', \etc\
When aligning names from two sets, the number of matches is upper\hyp bounded by the size of the smaller set, so we iterate over the smaller set and, for each name, find the optimal match in the larger set.
We only keep a match if its cosine similarity is above a threshold $\theta$ and if the best match has a much higher similarity than the second best match (to rule out ambiguities), by requiring a gap in similarities of at least $\delta$.
If this greedy procedure pairs the same element from the larger set with more than one element from the smaller set, we discard all these pairs.%
\footnote{
Alternatively, we could run a proper matching algorithm, such as the Hungarian algorithm, but as we aim to maximize precision, we opted for first being fully greedy and then generously discarding all potentially bad matches.
}

As mentioned, we use the same procedure for matching breweries and then for matching beers within breweries.
When matching breweries, we additionally require an exact match of locations (states for U.S.\ breweries, and countries for others).
When matching beers, we additionally require an exact match in alcohol by volume; also, before computing cosine similarities, we first remove from each beer name all tokens that also appear in the brewery name, as sometimes the name of the beer contains the brewery name in one dataset, but not in the other (\eg, ``Ingobr\"au Meistersud'' \vs just ``Meistersud'').

\subsection{Matched dataset}
\label{sec:matched_dataset}

The above algorithm has two parameters, $\theta$ and $\delta$.
We find $\theta=0.8$ and $\delta=0.3$ to work well in practice, as shown by an evaluation in which we inspected 500 matched brewery pairs and 500 matched beer pairs and, based on this ground truth, estimate precision as 99.6\% for matched breweries (2 of the 500 inspected matches were wrong) and 100\% for matched beers.

\begin{table}[tb]
\caption{Dataset size after matching.}
\vspace{-3mm}
\label{tbl:dataset size after matching}
\begin{tabular}{l|r|r|r|r}
& \multicolumn{4}{c}{Minimum number of ratings per beer} \\\cline{2-5}
& 0 & 5 & 10 & 20 \\\hline
Breweries & 6,084 & 2,561 & 1,711 & 1,079 \\
Beers & 45,640 & 12,890 & 7,424 & 4,051 \\
Ratings on BA & 955,968 & 873,944 & 812,070 & 732,165 \\
Ratings on RB & 1,020,638 & 761,496 & 650,642 & 542,961
\end{tabular}
\vspace{-3mm}
\end{table}


The size of the matched dataset is summarized in \Tabref{tbl:dataset size after matching}.
In our result analysis (\Secref{sec:results}), we restrict ourselves to beers with at least a minimum number of ratings, so the table lists sizes for various values of this threshold.
Although matching reduces the dataset by a lot, we are still left with tens of thousands of beers from thousands of breweries, with close to a million ratings on each site.

Matching ensures that each remaining product has been rated on each of the two websites.
As explained in \Secref{sec:steps} (step~2), each beer falls into one of nine \ptg{}s (HH, HM, HL, MH, \etc).
\Tabref{tbl:ptg counts} displays the size of all groups (for beers with at least five ratings on each site, as the bulk of our analysis will be conducted on this set).
We observe that the groups are rather well balanced ``out of the box'' (\eg, 1,210 beers in HM \vs 1,213 in MH, \etc), even before balancing them explicitly (step~3 of \Secref{sec:steps}).

\begin{table}[tb]
    \centering
    \caption{Number of beers per \ptg (Sec.\ \ref{sec:steps}, step 2) after matching, before balancing (\Secref{sec:steps}, step 3).
}
    \vspace{-3mm}
    \label{tbl:ptg counts}
    \begin{tabular}{rc|ccc}
        \multicolumn{2}{c}{} & \multicolumn{3}{c}{\textbf{BA}} \\ 
        \multicolumn{1}{c}{}& \multicolumn{1}{c}{}& H & M & L \\ 
        \hhline{~~|-|-|-|}
        \multirow{3}{*}{\textbf{RB}} & H & 585 & 1,213 & 116 \\
        & M & 1,210 & 6,593 & 1,242 \\
        & L & 138 & 1,225 & 568 
    \end{tabular}
    \vspace{-5mm}
\end{table}

We recomputed and inspected the rating histograms (\cf \Figref{fig:raw rating hist}) on the subset of beers in the matched sample only and found them essentially indistinguishable from the full sample shown in \Figref{fig:raw rating hist}.
This implies that the vastly different rating distributions of the two sites are not caused by users on one of the sites having a systematic preference for rating inherently better or worse beers (\eg, certain rating sites might see themselves as ``bashing sites'');
rather, the difference must stem from different scoring standards.

\begin{figure*}[t]
    \vspace{-3mm}
    \centering
    \subfigure[Histograms of ratings (std.) after matching]{
        \includegraphics[width=0.31\textwidth]{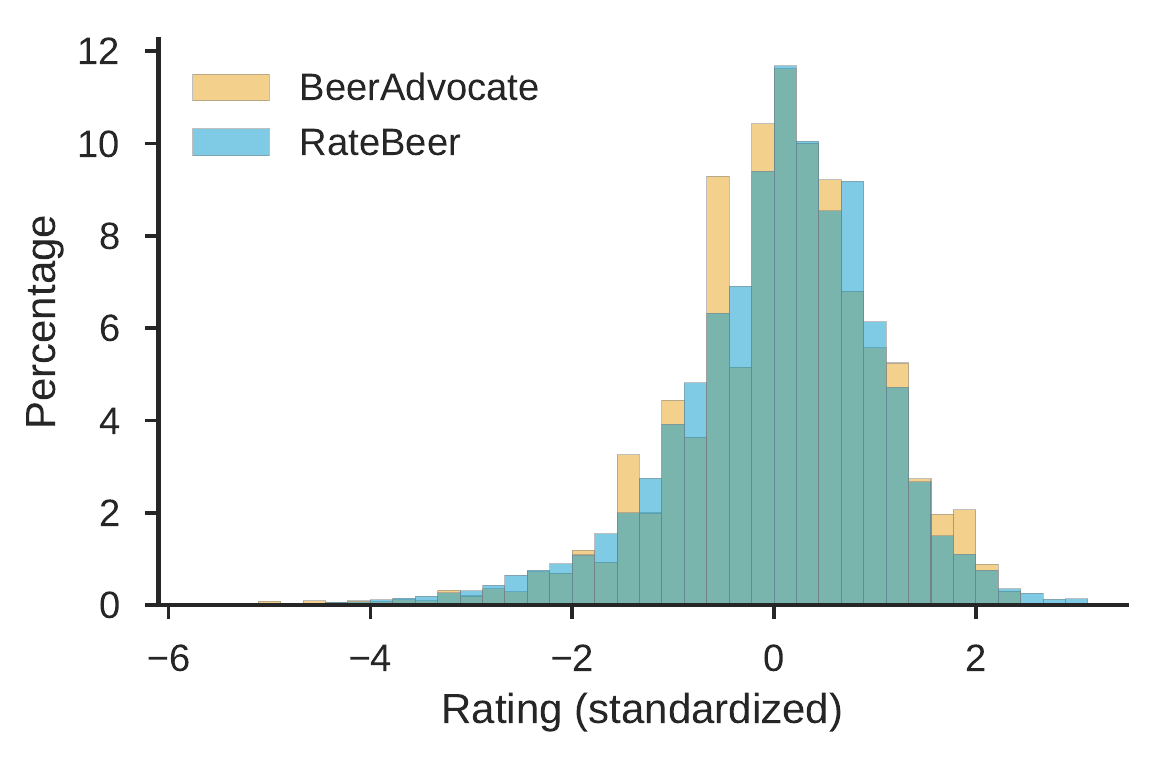}
        \label{fig:standardized matched rating hist}
    }
    \subfigure[Average ratings (std.) per beer after matching]{
        \includegraphics[width=0.27\textwidth]{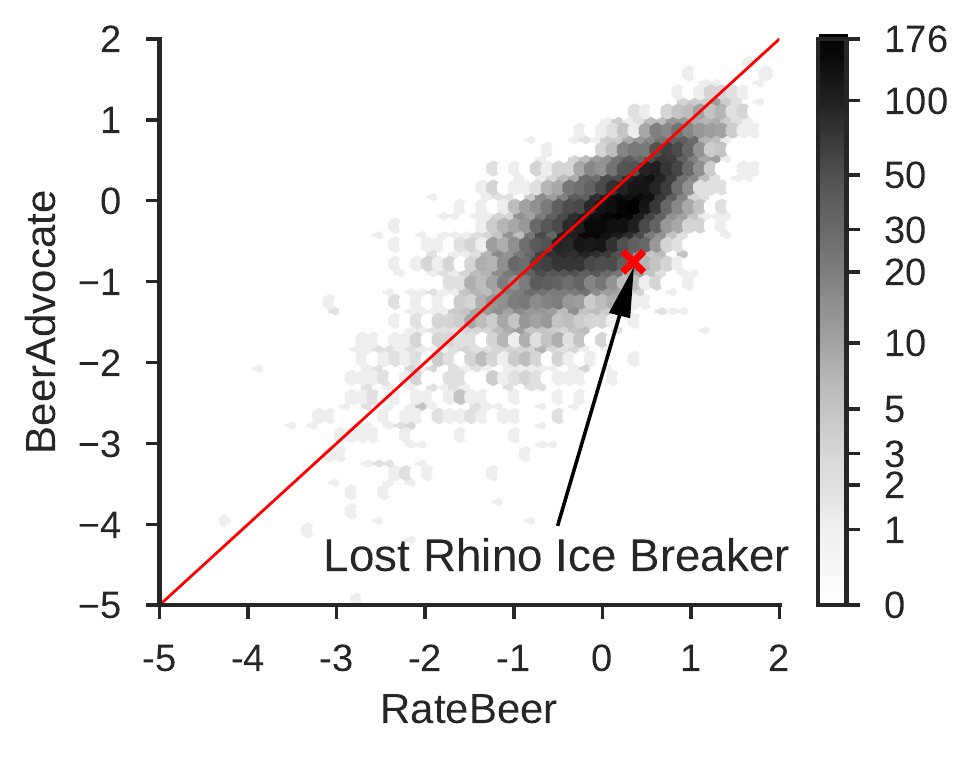}
        \label{fig:hexhist}
    }
    \vspace{-4mm}
    \caption{
    Standardized ratings after matching.
    Red cross in (b) marks example from \Figref{fig:example}; gray tones capture number of beers per bin.
    Although overall rating distributions are identical, many individual beers are rated differently across sites.
    }
    \label{fig:standardized and matched}
    \vspace{-3mm}
 \end{figure*}

To make scores comparable across sites, we thus standardize all ratings as described in \Secref{sec:basic analysis}, by subtracting the annual mean and dividing by the annual standard deviation. As seen in \Figref{fig:standardized matched rating hist}, the two sites' rating histograms are entirely overlapping after standardization, so ratings may now be compared across sites.

It is important to note that, whereas the distribution of standardized ratings is identical for the two sites (\Figref{fig:standardized matched rating hist}), there are numerous individual products with vastly different ratings on \BA \vs \RB.
To emphasize this point, \Figref{fig:hexhist} contains a scatter plot of ratings on \BA \vs ratings on \RB, where each beer is summarized by its average standardized rating on each site.
The fact that the point cloud disperses widely off the diagonal clearly shows that many beers are perceived differently on the two sites.
Viewed this way, the purpose of our natural experiment is to determine how beers such as \textit{Lost Rhino Ice Breaker} from the introduction wind up in the fringe of the point cloud of \Figref{fig:hexhist} (where \textit{Lost Rhino Ice Breaker} is marked as a red cross)---by virtue of herding or by sheer good or bad luck.

\subsection{Validity of matched sample}
\label{sec:validity}

Drawing correct conclusions from our observational study requires the assumptions laid out in \Secref{sec:assumptions}. The purpose of this section is to show that these assumptions, in particular external and internal validity, are empirically met by the matched beer rating dataset.

\xhdr{External validity}
Favoring precision over recall when matching (\Secref{sec:matching_algorithm}) comes at the expense of losing many matches in which we are less confident, which may introduce selection bias and can potentially impair the external validity of our results:
if the sample we study is fundamentally different from the overall population, our conclusions might not generalize from the former to the latter.

First recall from \Secref{sec:basic analysis} that there are some significant differences between the two sites:
\BA is more U.S.\hyp centric in terms of products and users, and it is also smaller in terms of the number of beers rated (\Tabref{tbl:dataset size}).
Since the number of matched beers is upper\hyp bounded by the number of beers in the smaller dataset, the best we could hope to do is match all beers in \BA to their corresponding beers in \RB.
This would preserve the original data distribution in \BA and skew \RB's distribution to match it.

\begin{figure*}[t]
    \vspace{-1mm}
    \centering
    \subfigure[Ratings]{
        \includegraphics[width=0.31\textwidth]{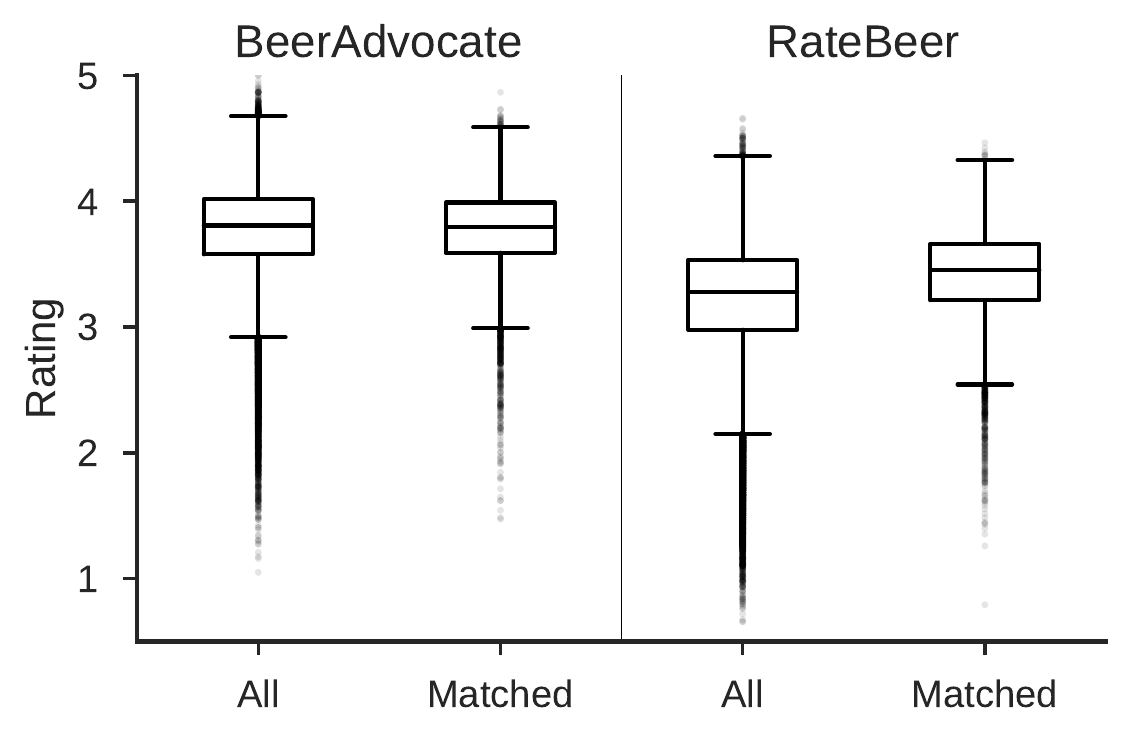}
        \label{fig:boxplot ratings}
    }
    \subfigure[Numbers of ratings per beer]{
        \includegraphics[width=0.31\textwidth]{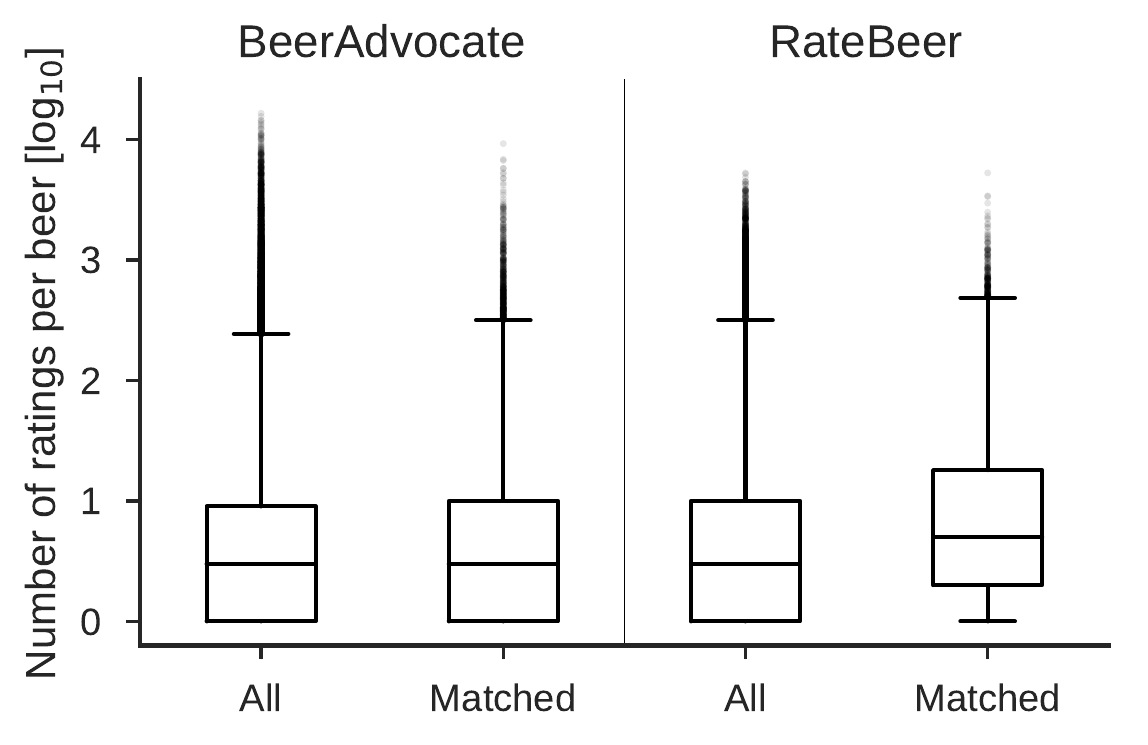}
        \label{fig:boxplot num ratings per beer}
    }
    \subfigure[Numbers of beers per brewery]{
        \includegraphics[width=0.31\textwidth]{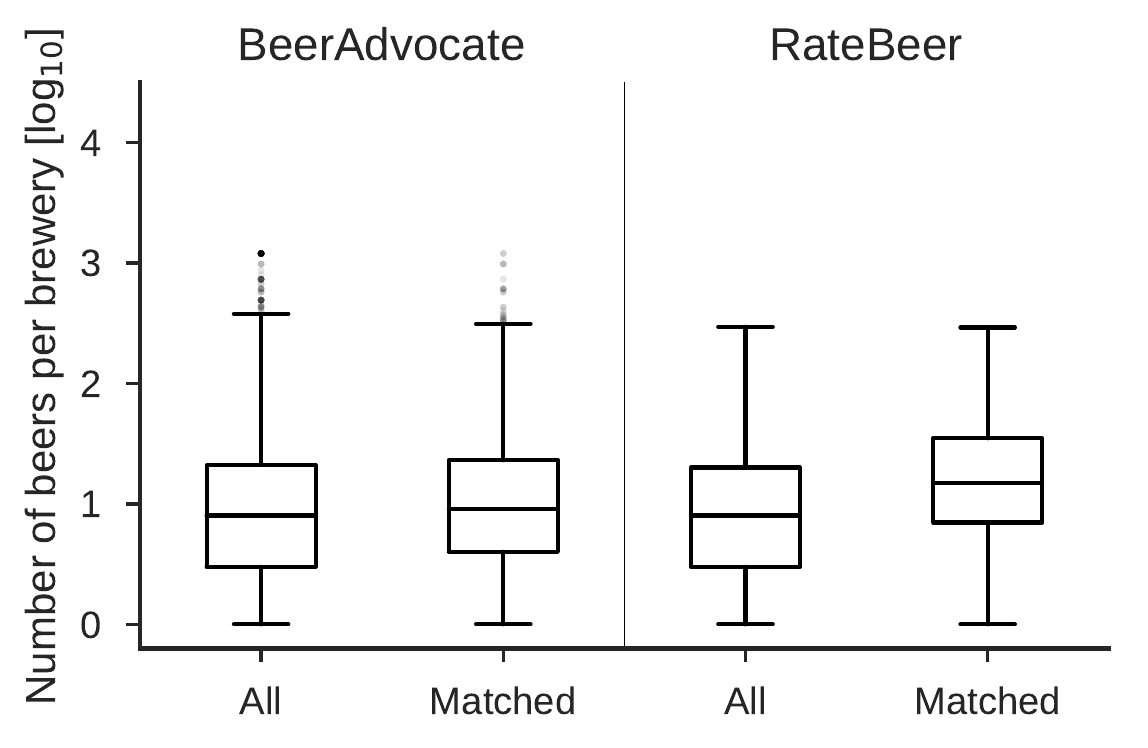}
        \label{fig:boxplot num beers per brewery}
    }
    \vspace{-4mm}
    \caption{
    Comparison of three dataset properties before \vs after matching.}
    \label{fig:boxplots pre vs post matching}
    \vspace{-3mm}
\end{figure*}

\Figref{fig:boxplots pre vs post matching} and \Tabref{tbl:brewery locations} show that, even though the smaller \BA is not a subset of \RB, matching still produces a dataset very similar to \BA.
\Figref{fig:boxplots pre vs post matching} inspects three exemplary properties (mean average beer rating, number of ratings per beer, and number of beers per brewery) of the data before (left box in each pair) \vs after matching (right box in each pair), for both \BA (left panel in each figure) and \RB (right panel in each figure).
We observe that matching does not noticeably alter \BA's distributions, whereas \RB's do change.

\Tabref{tbl:brewery locations} lists the most common countries of origin for breweries present in each dataset before matching, as well as in the matched dataset.
We make two observations.
First, the distribution over countries is similar in both datasets even before matching, with the exception that \BA contains a much larger fraction of U.S.\ breweries.
Second, matching mimics the distribution of the smaller dataset, \BA, more closely than that of \RB.
We also compared the style distributions before and after matching, with the same result that the matched dataset mirrors \BA's distribution closely
(in decreasing order of frequency:
American IPA 10.8\% before, \vs 12.1\% after, matching;
American Pale Ale 6.2\% \vs 6.8\%;
Saison\slash Farmhouse Ale 5.0\% \vs 5.9\%, \etc).

We conclude that our matched sample is unbiased with respect to \BA (as explained above, the best we could hope for), such that the conclusions we draw can at the very least be generalized to all of \BA.
Note that this is a conservative statement; we have seen no explicit indications why our conclusions should not also hold on all of \RB.

\begin{table}[tb]
  \caption{Brewery locations before and after matching (Sec.\ \ref{sec:steps}, step 1), ordered by percentage after matching.}
    \centering
    \begin{tabular}{l|c|c|c}
         & \multicolumn{2}{c|}{Unmatched} & Matched \\ 
        & BA & RB & \\ \hline
        United States & 44.4\% & 28.6\% & 47.8\% \\
        Germany & 8.5\% & 8.3\% & 6.4\% \\
        England & 6.1\% & 8.8\% & 5.8\% \\
        Canada & 5.1\% & 3.7\% & 4.9\% \\
        Italy & 2.2\% & 4.3\% & 2.7\% \\
        Belgium & 2.0\% & 1.9\% & 2.5\% \\
        France & 2.4\% & 3.5\% & 2.3\% \\
        Spain & 1.9\% & 3.2\% & 2.3\% \\
        Australia & 2.4\% & 2.3\% & 2.2\% \\
        Netherlands & 1.5\% & 2.1\% & 2.1\% \\
    \end{tabular}
  \label{tbl:brewery locations}
  \vspace{-4mm}
\end{table}












\xhdr{Internal validity}
As argued in \Secref{sec:assumptions}, we need to show that the treatment assignment $T$ (the \fr a beer receives) is independent of the rating site $S$ and the rated product $P$.
Although, as discussed there, we have $T \indep S$ and $T \indep P$ by construction, these do not automatically imply $T \indep (S,P)$.
For instance, it is in principle possible (though not likely) that users on site $S_1$ love all pale beers and hate all dark beers, while users on $S_2$ love all dark beers and hate all pale beers.
This would entail that all pale beers would see both a high treatment (\fr) and a high outcome (subsequent reviews) on $S_1$;
and that all dark beers would see both a high treatment and a high outcome on $S_2$.
Here, a correlation between treatment and outcome would not be causal, but due to the confound of site\hyp specific preferences.
We therefore need to check empirically that the distribution of treatment assignments (\ie, the probability of receiving a higher \fr) is approximately equal for all combinations of site and product properties.
Notice that only properties available \textit{before} treatment should be taken into account here, as all other properties might be consequences, rather than causes, of the treatment.
This precludes us, \eg, from considering ratings received by the respective beer.

Inspecting treatment probabilities for all beer properties on each site would be elusive, especially given our limited dataset size.
But we argue that the most likely confounds would be captured by beer style and producer country:
controlling for style also roughly fixes the most salient properties of a beer, such as bitterness, color, alcohol content, \etc;
and controlling for producer country accounts for the fact that \BA users are more likely American and might therefore be biased toward (or against) American beers.

The numbers are presented in \Tabref{tbl:internal validity} for the most frequent styles and countries.
There is one table per group of interest (HM, ML, HL).
For each combination of beer property (style or country) and site, we list the number of beers with the higher treatment on that site and the implied probability of receiving the higher treatment.
As there are two treatments in each group, perfect independence would yield probabilities of 50\% everywhere.

While achieving such an exact balance is infeasible with our limited dataset, \Tabref{tbl:internal validity} shows that we come rather close;
in particular, we achieve treatment probabilities close to 50\% for the most frequent styles and countries (which have the biggest impact).

These numbers mean that our cross-site product matching results in a treatment assignment that is (approximately) independent of site and product properties (\cf \Figref{fig:bayesian_networks_good_natexp}).
Though not randomized by us as the researchers, treatment assignment is mostly haphazard, such that we may indeed speak of a natural experiment.

\begin{table*}[tb]
    \caption{
    Counts (\#) and probabilities (Pr) of higher treatment for three \ptg{}s, both sites (BA, RB), and top beer styles and brewery countries.
    Most values being similar for BA and RB supports internal validity of our study (\Secref{sec:validity}).
    }
    \vspace{-3mm}
    \centering
    {\footnotesize
    \begin{tabular}{c|c|c}
    \hspace{-4mm}
    \begin{tabular}{l|cc|cc}
         \multicolumn{5}{c}{\textbf{HM}} \\
         \multicolumn{1}{c}{} & \multicolumn{2}{c}{\#(H)} & \multicolumn{2}{c}{Pr(H)} \\
         \hline
         Style & BA & RB & BA & RB \\\hline
         Amer.\ IPA & 105 & 88 & 0.54 & 0.46 \\
         Amer.\ Double/Imp.\ IPA & 100 & 81 & 0.55 & 0.45 \\
         Amer.\ Pale Ale & 32 & 32 & 0.50 & 0.50 \\
         Amer.\ Wild Ale & 27 & 36 & 0.43 & 0.57 \\
         Saison/Farmhouse Ale & 33 & 30 & 0.52 & 0.48 \\
         Amer.\ Double/Imp.\ Stout & 22 & 36 & 0.38 & 0.62 \\
         Amer.\ Black Ale & 15 & 20 & 0.43 & 0.57 \\
         Amer.\ Porter & 14 & 18 & 0.44 & 0.56 \\
         Russian Imperial Stout & 13 & 13 & 0.50 & 0.50 \\
         American Stout & 10 & 14 & 0.42 & 0.58 \\
         \hline \hline
         Country & BA & RB & BA & RB \\\hline
         United States & 544 & 493 & 0.52 & 0.48 \\
         Canada & 24 & 37 & 0.39 & 0.61 \\
         Belgium & 30 & 31 & 0.49 & 0.51 \\
         England & 7 & 13 & 0.35 & 0.65 \\
         Australia & 4 & 14 & 0.22 & 0.78 \\
         Germany & 10 & 7 & 0.59 & 0.41 \\
         Sweden & 10 & 6 & 0.62 & 0.38 \\
         Italy & 3 & 8 & 0.27 & 0.73 \\
         Denmark & 4 & 7 & 0.36 & 0.64 \\
         New Zealand & 0 & 9 & 0.00 & 1.00 \\
         \hline
    \end{tabular}
&
    \begin{tabular}{l|cc|cc}
         \multicolumn{5}{c}{\textbf{ML}} \\
         \multicolumn{1}{c}{} & \multicolumn{2}{c}{\#(M)} & \multicolumn{2}{c}{Pr(M)} \\
         \hline
         Style & BA & RB & BA & RB \\\hline
         Amer.\ IPA & 64 & 64 & 0.50 & 0.50 \\
         Amer.\ Pale Ale & 36 & 44 & 0.45 & 0.55 \\
         Fruit/Vegetable Beer & 30 & 25 & 0.55 & 0.45 \\
         Amer.\ Amber/Red Ale & 32 & 20 & 0.61 & 0.39 \\
         Amer.\ Double/Imp.\ IPA & 28 & 18 & 0.61 & 0.39 \\
         Saison/Farmhouse Ale & 20 & 26 & 0.44 & 0.56 \\
         Amer.\ Blonde Ale & 28 & 13 & 0.68 & 0.32 \\
         Amer.\ Porter & 14 & 24 & 0.37 & 0.63 \\
         Amer.\ Pale Wheat Ale & 18 & 19 & 0.49 & 0.51 \\
         German Pilsener & 14 & 21 & 0.40 & 0.60 \\
         \hline \hline
         Country & BA & RB & BA & RB \\\hline
         United States & 445 & 449 & 0.50 & 0.50 \\
         Canada & 84 & 53 & 0.61 & 0.39 \\
         Belgium & 28 & 36 & 0.44 & 0.56 \\
         Germany & 20 & 21 & 0.49 & 0.51 \\
         Australia & 20 & 19 & 0.51 & 0.49 \\
         England & 9 & 19 & 0.32 & 0.68 \\
         Netherlands & 7 & 12 & 0.37 & 0.63 \\
         Italy & 4 & 11 & 0.27 & 0.73 \\
         Scotland & 5 & 6 & 0.46 & 0.54 \\
         New Zealand & 3 & 5 & 0.38 & 0.62 \\
         \hline
    \end{tabular}
&
    \begin{tabular}{l|cc|cc}
         \multicolumn{5}{c}{\textbf{HL}} \\
         \multicolumn{1}{c}{} & \multicolumn{2}{c}{\#(H)} & \multicolumn{2}{c}{Pr(H)} \\
         \hline
         Style & BA & RB & BA & RB \\\hline
         Amer.\ IPA & 7 & 5 & 0.58 & 0.42 \\
         Amer.\ Double/Imp.\ IPA & 8 & 2 & 0.80 & 0.20 \\
         Amer.\ Pale Ale & 3 & 3 & 0.50 & 0.50 \\
         Russian Imperial Stout & 0 & 5 & 0.00 & 1.00 \\
         Amer.\ Amber/Red Ale & 3 & 2 & 0.60 & 0.40 \\
         Amer.\ Wild Ale & 4 & 1 & 0.80 & 0.20 \\
         Amer.\ Barleywine & 2 & 3 & 0.40 & 0.60 \\
         Amer.\ Blonde Ale & 1 & 3 & 0.25 & 0.75 \\
         Amer.\ Porter & 2 & 2 & 0.50 & 0.50 \\
         Amer.\ Black Ale & 0 & 4 & 0.00 & 1.00 \\
         \hline \hline
         Country & BA & RB & BA & RB \\\hline
         United States & 46 & 38 & 0.55 & 0.45 \\
         Canada & 7 & 7 & 0.50 & 0.50 \\
         Germany & 3 & 6 & 0.33 & 0.67 \\
         Belgium & 5 & 2 & 0.71 & 0.29 \\
         Australia & 2 & 4 & 0.33 & 0.67 \\
         Switzerland & 1 & 2 & 0.33 & 0.67 \\
         Denmark & 0 & 2 & 0.00 & 1.00 \\
         Austria & 1 & 1 & 0.50 & 0.50 \\
         Netherlands & 1 & 1 & 0.50 & 0.50 \\
         New Zealand & 1 & 1 & 0.50 & 0.50 \\
         \hline
    \end{tabular}
    \end{tabular}
}
    \label{tbl:internal validity}
\end{table*}

\section{Results}
\label{sec:results}
Now that we have introduced our matched dataset and established that correlations between first and later ratings are highly likely to be causal, we proceed to analyzing the dataset with respect to such correlations, to estimate the effect of first on later ratings.

After processing the data as described in \Secref{sec:steps} (steps 1--4), there are 3 \ptg{}s of interest: HL, HM, ML.
We estimate the effect of a higher (H for HL\slash HM; M for ML) \vs a lower (L for HL\slash ML; M for HM) \fr separately for each of them.

Recall that each group contains the same beer twice, once on \BA, once on \RB; and once with a higher, once with a lower, \fr.
As established in \Secref{sec:validity}, on which site a beer receives the higher \fr is essentially haphazard.
Therefore, in the presence of herding, a beer will on average receive higher subsequent ratings on the site on which it happened to receive the higher \fr, and lower subsequent ratings on the other site.
In the absence of herding, subsequent ratings will be indistinguishable between the two sites on average.
In other words, we have herding if and only if first and subsequent ratings are correlated.

\begin{figure*}[t]
    \vspace{-3mm}
    \centering
    \subfigure[Extreme \ptg{}s]{
        \includegraphics[width=0.25\textwidth]{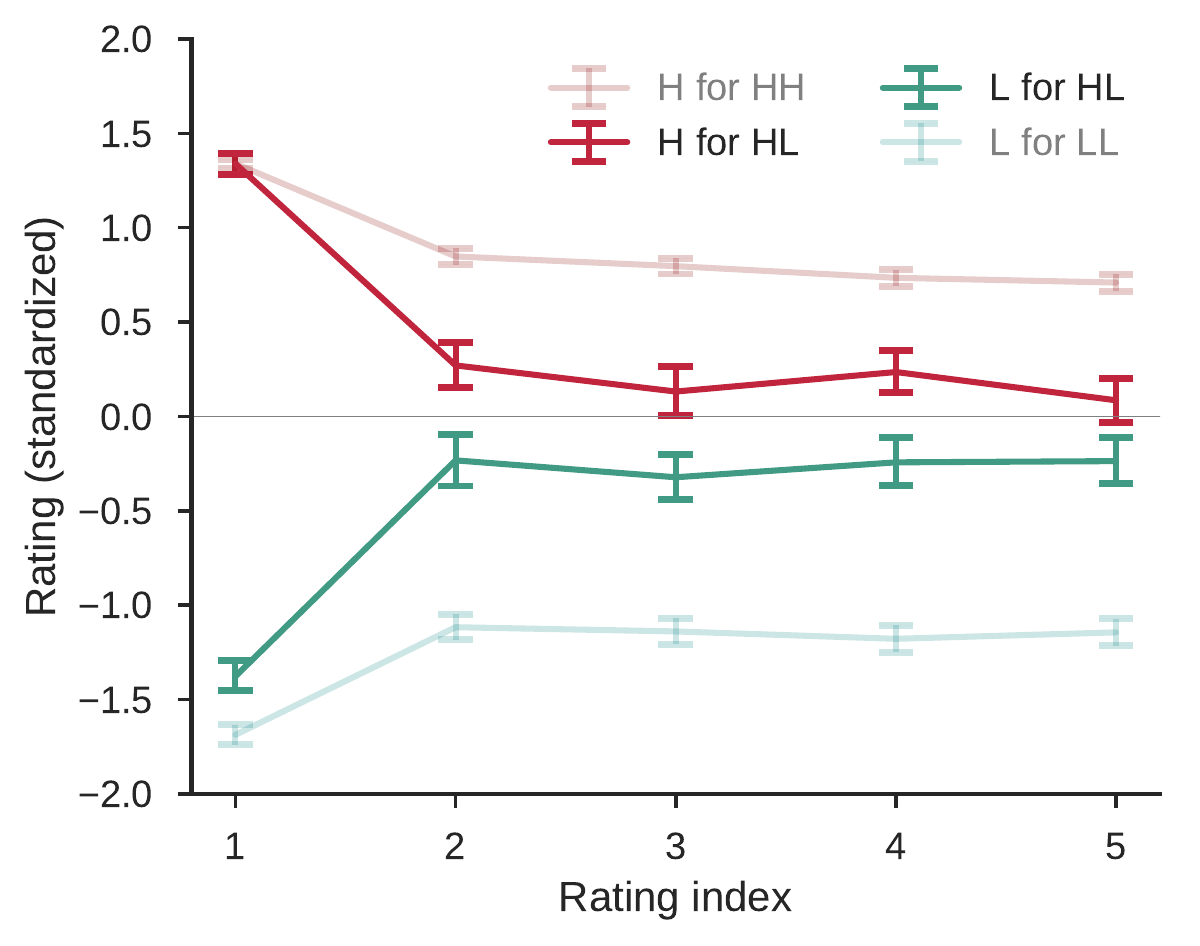}
        \label{fig:herding_extreme}
    }
    \subfigure[Less extreme \ptg{}s]{
        \includegraphics[width=0.25\textwidth]{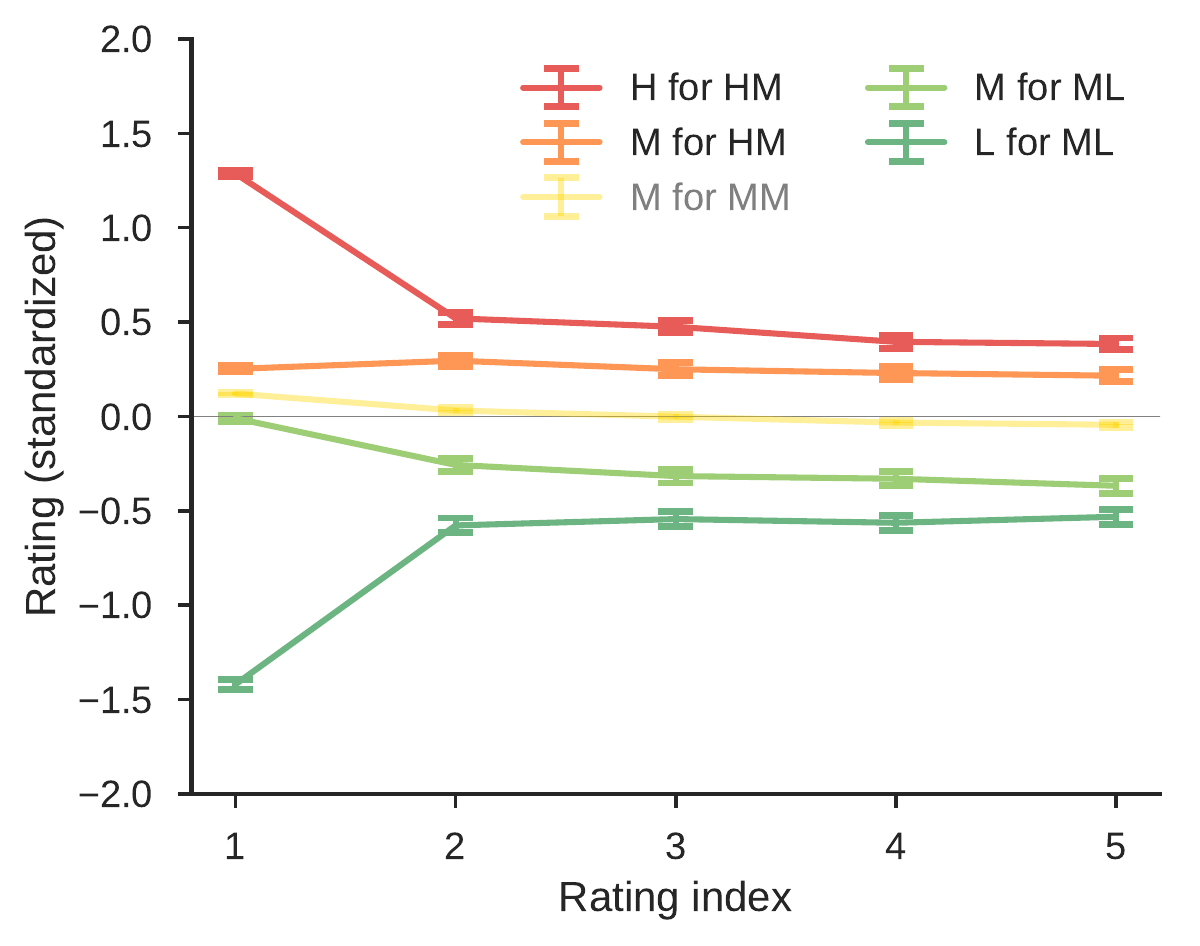}
        \label{fig:herding_medium}
    }
    \subfigure[Long-term average ratings]{
        \includegraphics[width=0.45\textwidth]{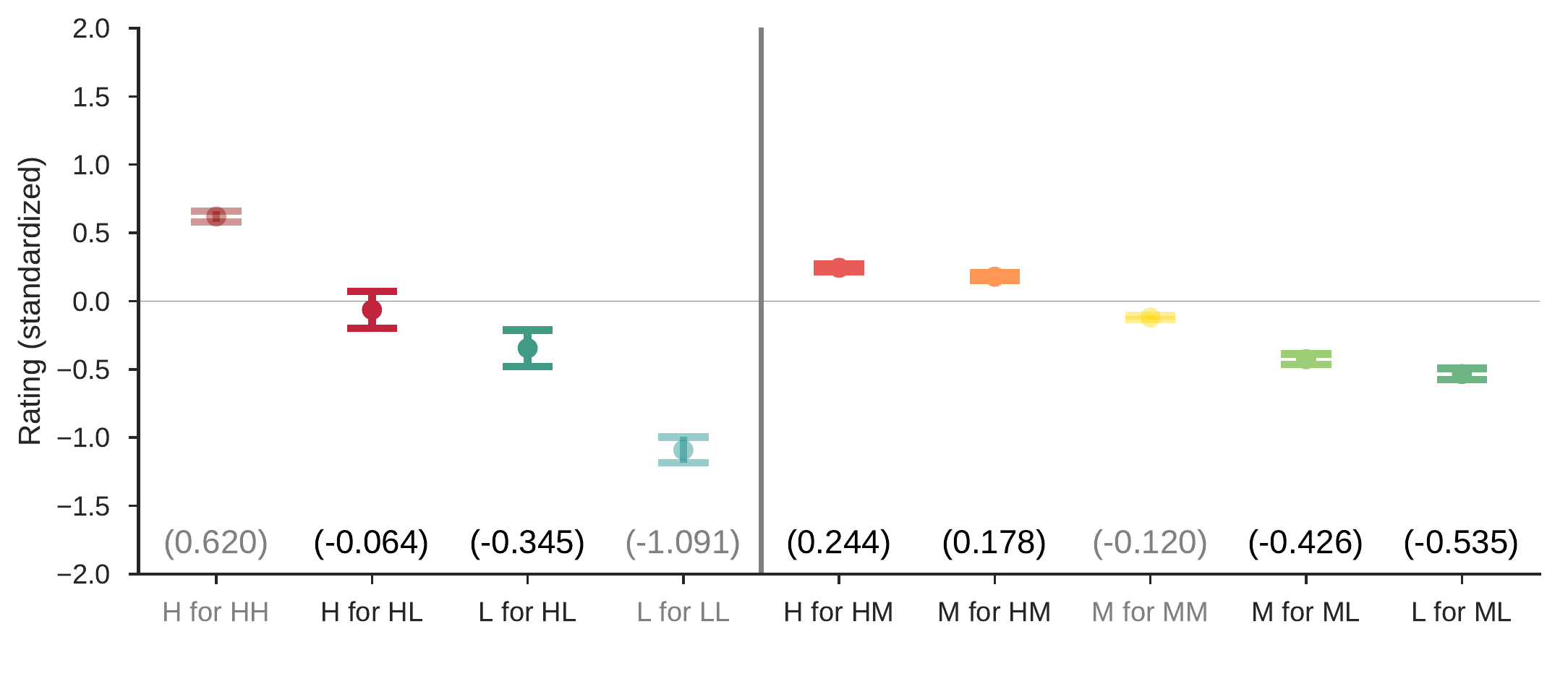}
        \label{fig:herding_longterm}
        \vspace*{2cm}
    }
    \vspace{-4mm}
    \caption{
    Herding effects in beer ratings;
    (a--b) identical beer receives significantly higher (lower) subsequent ratings if \fr was higher (lower);
    (c) herding effects last long: even after 20 or more ratings, product average ratings are significantly higher (lower) for beers with higher (lower) \fr{}s.
    Ratings are standardized; error bars show 95\% confidence intervals.
    }
    \label{fig:herding results zscores}
    \vspace{-3mm}
\end{figure*}

In this light, \Figref{fig:herding_extreme} provides a clear indication of herding.
The dark curves in the center correspond to the \ptg HL.
The dark red (green) curve summarizes ratings received on the site with a \fr of H (L); the horizontal axis shows the rating index $i=1,\dots,5$, and the vertical axis, the $i$-th standardized rating on the respective site, averaged over all beers in the HL group.
The plot includes only beers with at least five ratings, so the same set of beers contributes to all indices $i$.
We also emphasize that the dark red and green curves are computed on exactly the same set of beers, just with a different treatment per curve.
Thus, we conclude from \Figref{fig:herding_extreme} that the same beer's second rating is about half a standard deviation higher if the \fr was H, \vs if it was L.

The same effect can be observed for the less extreme \ptg{}s (HM and ML; \Figref{fig:herding_medium}), but with less extreme differences, as expected: here, a different \fr translates into a second\hyp rating difference of about a quarter standard deviation.
\Figref{fig:herding results zscores}(a--b) also show that the impact of herding extends beyond the second rating; the effect size is roughly constant on any of the first five ratings.
It is therefore interesting to ask whether the herding effect lingers indefinitely or is ultimately overridden by the inherent quality of the respective beer.
To address this question, we consider only beers with a substantial number of ratings (at least 20) and compare their long-term averages (based on all ratings received up until the datasets were crawled) for different \fr{}s.
The results of this analysis, plotted in \Figref{fig:herding_longterm}, are clear:
even after 20 or more ratings, a high \fr (H for HL) entails a rating on average 0.28 standard deviations higher than what we would see after a low \fr (L for HL).
The effect is again less extreme for less extreme \fr differences (HM, ML), but it is still noticeable (and with non\hyp overlapping 95\% confidence intervals).


The plots of \Figref{fig:herding results zscores} were computed after aggregating symmetric \ptg{}s, such that, \eg, beers that received H on \BA and L on \RB were grouped together with beers that received L on \BA and H on \RB (\cf step~4 in \Secref{sec:steps}).
As a sanity check, we also investigate treatment differences in the non\hyp aggregated groups (\Tabref{tbl:herding effect directional}), concluding that a higher \fr entails higher subsequent ratings regardless of the site on which the higher \fr occurred.

\begin{table*}[h]
    \renewcommand{\arraystretch}{1.2}
    \caption{
    Standardized 5th rating (with 95\% confidence intervals) for disaggregated \ptg{}s, \ie, without step~4 of \Secref{sec:steps};
    \eg, cell (L, H on RB) contains mean 5th rating on site with \fr L when \fr H happened on RB.
    }
    \vspace{-3mm}
    \centering
{\footnotesize
\begin{tabular}{ccc}
    \begin{tabular}{c|cc}
    \multicolumn{1}{c}{} & H on BA & H on RB \\ \cline{2-3}
    H & -0.032 [-0.193, 0.123] & 0.226 [0.045, 0.393] \\
    L & -0.392 [-0.572, -0.231] & -0.066 [-0.220, 0.099] \\
    \end{tabular}
&
    \begin{tabular}{c|cc}
    \multicolumn{1}{c}{} & H on BA & H on RB \\ \cline{2-3}
    H & 0.270 [0.225, 0.314] & 0.498 [0.460, 0.542] \\
    M & 0.057 [0.006, 0.107] & 0.376 [0.334, 0.423] \\
    \end{tabular}
&
    \begin{tabular}{c|cc}
    \multicolumn{1}{c}{} & M on BA & M on RB \\ \cline{2-3}
    M & -0.506 [-0.563, -0.450] & -0.229 [-0.279, -0.176] \\
    L & -0.657 [-0.721, -0.596] & -0.414 [-0.467, -0.363] \\
    \end{tabular}
    \vspace{-3mm}
\end{tabular}
}
    \label{tbl:herding effect directional}
\end{table*}

\Figref{fig:herding results zscores} and \Tabref{tbl:herding effect directional} directly capture
(standardized) rating scores. Plain scores are, however, not all that matters; rating sites tend to
also rely heavily on rankings, \eg, when making recommendations to users.
Therefore, we also repeat our analysis by measuring outcomes in terms of ranks (normalized to lie between 0 and 1), rather than scores, as follows (we explain our setup for the HL group; it is analogous for the other groups).
For each rating index $i=1,\dots,5$, we rank all beers in the group by their $i$-th rating, resulting in one ranking for each of the two sites.
Then, we compute, for each beer, its mean rank on the site where it received H as the \fr, as well as on the site where it received L.
Comparing the mean for H with the mean for L, we observe that normalized ranks with respect to second ratings are 61\% on average on the H site, and only 43\% on the L site.
We consider this normalized\hyp rank difference of 18\% rather substantial.
(Even when considering fifth, rather than second, ratings, we still measure a difference of 10\%.)

\section{Discussion and related work}
\label{sec:discussion}
\xhdr{Summary of results}
Our results show clear evidence of herding in the ratings collected by the two most prominent online beer\hyp rating platforms.
We find that a very high \fr leads to the following five ratings being about half a standard deviation higher, compared to a situation in which the exact same beer receives a very low \fr (\Figref{fig:herding results zscores}(a--b)).
These differences in absolute scores also translate into large differences in ranking positions (18\% for the second, 10\% for the fifth, rating), which are essential for product recommendations.
The problem would be mitigated if the platform managed to swiftly ``forget'' the first reviews and to converge to the true, inherent quality of the beer being rated, but we find that this is not the case, with the effects of \fr{}s lingering until after the beer has received 20 or more reviews (\Figref{fig:herding_longterm}).

Whether the first review is positive or negative might come down to random factors such as if the sun was shining when the first reviewer tasted the beer, if they had a bad stomach, or if they had been in a fight with their husband, which may then kick off a domino effect with potentially severe consequences: since many users rely on rating sites to decide what to buy, randomness among the first reviews can tangibly affect the business of producers.

\xhdr{Implications for rating\hyp site design}
A simple idea to address this problem would be to hide all reviews of a product as long as it has received less than a minimum number of ratings. If, \eg, ratings are hidden until there are at least ten of them, this will mean that, effectively, the ten first ratings are independent of one another and not affected by herding.
Once the eleventh reviewer arrives, they will see an average rating that reflects the inherent quality of the product much more closely than any one single review.
As a consequence, even if the eleventh reviewer is biased by previous ratings, they will be biased by something much less haphazard.
Future work should verify this hypothesis in an A/B test.

\xhdr{Community overlap}
We point out that our conclusions hold even when information flows between the two sites via users active on both sites (such users exist in practice):
as we focus on products with divergent \fr{}s, a surmised dependence between the two sites (with respect to the products we study) would have to be one of deliberate anti\hyp herding, which would be hard to explain.

\xhdr{Prior work on herding}
Early work on human herding behavior (primarily from marketing and economics) was inspired by work from biology on the behavior of animal herds \cite{hamilton1971geometry},
which possibly explains why some of the earliest empirical studies of human herding considered farmers \cite{yoav1973process} and investment bankers \cite{scharfstein1990herd}.

Due to data scarcity, much early work was theoretical \cite{banerjee1992simple}, but with the rise of the Web, empirical studies have become more feasible.
The strongest evidence, naturally, comes from experimental studies.
Prominently, Muchnik \etal \cite{muchnik2013social} inserted random \fr{}s into a news\hyp story website and studied how users reacted to these treatments.
Interestingly, while they found that both up- and down-votes skew later votes, the effects of down-votes were offset by social correction in their case, \ie, by benign users over\hyp compensating for down-votes with subsequent up-votes.
In our case, even if social correction should occur, it certainly does not override the negative influence of haphazard early ratings (\Figref{fig:herding results zscores}).
A follow\hyp up experiment \cite{glenski2017rating} yielded evidence for herding on the social bookmarking site Reddit, but without evidence for social correction.

Experiments are powerful tools, but they are expensive to run and involve random manipulations that raise ethical challenges, which, \eg, kept Muchnik \etal \cite{muchnik2013social} from disclosing the site on which they had operated.
Simulations \cite{wang2014amazon} and observational studies \cite{chevalier2006effect,duan2008online,lee2015follow} can serve as an alternative, but circumventing the problems we have mentioned in the introduction tends to require complex modeling assumptions and ways to control for confounds.

We, on the contrary, propose a methodology based on natural experiments, which, although also observational, eliminates the need for explicitly controlling for confounds by leveraging a situation where treatment assignment is haphazard.
Our approach is inspired by the method of \textit{double pair comparison} \cite{evans1986effectiveness,evans1986double}, which was first applied to study the effectiveness of car safety belts (a concise summary of the study is given by Rosenbaum \cite[Sec.~1.4]{rosenbaum2002observational}).

\xhdr{Applicability of our method}
Being cheap and not interfering with the systems being studied are clear advantages of observational studies, especially because different settings may be affected by herding in different ways (\cf the above case of social correction on news stories \cite{muchnik2013social} \vs Reddit posts \cite{glenski2017rating}), such that we should study a variety of cases.
Luckily, our method applies very generally (\Secref{sec:methodology}).
We simply require a set of products rated on two separate websites and alignable across the two sites.
We emphasize again the importance of verifying the validity of each setting before analyzing results.
In particular, we need to ascertain that the matched sample of products is unbiased with respect to the set of all samples (external validity),
and that matching products across sites indeed results in \fr{}s (treatment assignment) being independent of product and site properties (internal validity).

When assessing validity, we are limited to observed product features.
In particular, we argued that the style and country of a beer are the primary potential confounds, as they capture most other conceivable confounds, observed or unobserved (\Secref{sec:validity}).
Despite this extrinsic argument, we stress that one can never fully rule out unobserved confounds, something researchers should be aware of when applying our method to other datasets.
When one does not have overwhelming extrinsic arguments supporting the independence of treatment from product and site properties (internal validity), one may perform a sensitivity analysis \cite{rosenbaum2002observational} to quantify how strongly the treatment would have to depend on such properties before we would alter our conclusions.


\xhdr{Future work}
We hope that researchers will adopt our method to study herding in further scenarios.
We believe that product ratings on Amazon constitute a particularly interesting case, as Amazon has sites in multiple languages (\eg, Amazon.com, Amazon.de, Amazon.fr), each with an independent rating system, yet covering overlapping subsets of a wide spectrum of products.
Also, as each product has a unique Amazon\hyp wide identifier, matching is trivial.

Our results raise several interesting questions:
Are certain users (\eg, newcomers) more susceptible to herding than others?
Can exposure to haphazard ratings lastingly alter a user's later behavior (rather than only a product's later ratings)?
And finally, given ratings for the same product from several websites, can we develop models for combining them into a more truthful aggregate score?





\xhdr{Acknowledgments}
We thank Julian McAuley and Carlos Castillo for thoughtful discussions.

\bibliographystyle{ACM-Reference-Format}

\begin{thebibliography}{18}


\ifx \showCODEN    \undefined \def \showCODEN     #1{\unskip}     \fi
\ifx \showDOI      \undefined \def \showDOI       #1{#1}\fi
\ifx \showISBNx    \undefined \def \showISBNx     #1{\unskip}     \fi
\ifx \showISBNxiii \undefined \def \showISBNxiii  #1{\unskip}     \fi
\ifx \showISSN     \undefined \def \showISSN      #1{\unskip}     \fi
\ifx \showLCCN     \undefined \def \showLCCN      #1{\unskip}     \fi
\ifx \shownote     \undefined \def \shownote      #1{#1}          \fi
\ifx \showarticletitle \undefined \def \showarticletitle #1{#1}   \fi
\ifx \showURL      \undefined \def \showURL       {\relax}        \fi
\providecommand\bibfield[2]{#2}
\providecommand\bibinfo[2]{#2}
\providecommand\natexlab[1]{#1}
\providecommand\showeprint[2][]{arXiv:#2}

\bibitem[\protect\citeauthoryear{Banerjee}{Banerjee}{1992}]%
        {banerjee1992simple}
\bibfield{author}{\bibinfo{person}{Abhijit~V Banerjee}.}
  \bibinfo{year}{1992}\natexlab{}.
\newblock \showarticletitle{A Simple Model of Herd Behavior}.
\newblock \bibinfo{journal}{{\em The Quarterly Journal of Economics\/}}
  \bibinfo{volume}{107}, \bibinfo{number}{3} (\bibinfo{year}{1992}),
  \bibinfo{pages}{797--817}.
\newblock


\bibitem[\protect\citeauthoryear{Chevalier and Mayzlin}{Chevalier and
  Mayzlin}{2006}]%
        {chevalier2006effect}
\bibfield{author}{\bibinfo{person}{Judith~A Chevalier} {and}
  \bibinfo{person}{Dina Mayzlin}.} \bibinfo{year}{2006}\natexlab{}.
\newblock \showarticletitle{The Effect of Word of Mouth on Sales: Online Book
  Reviews}.
\newblock \bibinfo{journal}{{\em Journal of Marketing Research\/}}
  \bibinfo{volume}{43}, \bibinfo{number}{3} (\bibinfo{year}{2006}),
  \bibinfo{pages}{345--354}.
\newblock


\bibitem[\protect\citeauthoryear{Duan, Gu, and Whinston}{Duan
  et~al\mbox{.}}{2008}]%
        {duan2008online}
\bibfield{author}{\bibinfo{person}{Wenjing Duan}, \bibinfo{person}{Bin Gu},
  {and} \bibinfo{person}{Andrew~B Whinston}.} \bibinfo{year}{2008}\natexlab{}.
\newblock \showarticletitle{Do Online Reviews Matter? An Empirical
  Investigation of Panel Data}.
\newblock \bibinfo{journal}{{\em Decision Support Systems\/}}
  \bibinfo{volume}{45}, \bibinfo{number}{4} (\bibinfo{year}{2008}),
  \bibinfo{pages}{1007--1016}.
\newblock


\bibitem[\protect\citeauthoryear{Evans}{Evans}{1986a}]%
        {evans1986double}
\bibfield{author}{\bibinfo{person}{Leonard Evans}.}
  \bibinfo{year}{1986}\natexlab{a}.
\newblock \showarticletitle{Double Pair Comparison: A New Method to Determine
  How Occupant Characteristics Affect Fatality Risk in Traffic Crashes}.
\newblock \bibinfo{journal}{{\em Accident Analysis \& Prevention\/}}
  \bibinfo{volume}{18}, \bibinfo{number}{3} (\bibinfo{year}{1986}),
  \bibinfo{pages}{217--227}.
\newblock


\bibitem[\protect\citeauthoryear{Evans}{Evans}{1986b}]%
        {evans1986effectiveness}
\bibfield{author}{\bibinfo{person}{Leonard Evans}.}
  \bibinfo{year}{1986}\natexlab{b}.
\newblock \showarticletitle{The Effectiveness of Safety Belts in Preventing
  Fatalities}.
\newblock \bibinfo{journal}{{\em Accident Analysis \& Prevention\/}}
  \bibinfo{volume}{18}, \bibinfo{number}{3} (\bibinfo{year}{1986}),
  \bibinfo{pages}{229--241}.
\newblock


\bibitem[\protect\citeauthoryear{Glenski and Weninger}{Glenski and
  Weninger}{2017}]%
        {glenski2017rating}
\bibfield{author}{\bibinfo{person}{Maria Glenski} {and} \bibinfo{person}{Tim
  Weninger}.} \bibinfo{year}{2017}\natexlab{}.
\newblock \showarticletitle{Rating Effects on Social News Posts and Comments}.
\newblock \bibinfo{journal}{{\em ACM Transactions on Intelligent Systems and
  Technology\/}} \bibinfo{volume}{8}, \bibinfo{number}{6}
  (\bibinfo{year}{2017}), \bibinfo{pages}{78}.
\newblock


\bibitem[\protect\citeauthoryear{Hamilton}{Hamilton}{1971}]%
        {hamilton1971geometry}
\bibfield{author}{\bibinfo{person}{William~D Hamilton}.}
  \bibinfo{year}{1971}\natexlab{}.
\newblock \showarticletitle{Geometry for the Selfish Herd}.
\newblock \bibinfo{journal}{{\em Journal of Theoretical Biology\/}}
  \bibinfo{volume}{31}, \bibinfo{number}{2} (\bibinfo{year}{1971}),
  \bibinfo{pages}{295--311}.
\newblock


\bibitem[\protect\citeauthoryear{Hu, Pavlou, and Zhang}{Hu
  et~al\mbox{.}}{2006}]%
        {hu2006can}
\bibfield{author}{\bibinfo{person}{Nan Hu}, \bibinfo{person}{Paul~A Pavlou},
  {and} \bibinfo{person}{Jennifer Zhang}.} \bibinfo{year}{2006}\natexlab{}.
\newblock \showarticletitle{Can Online Reviews Reveal a Product's True Quality?
  Empirical Findings and Analytical Modeling of Online Word-of-Mouth
  Communication}. In \bibinfo{booktitle}{{\em Proceedings of the 7th ACM
  Conference on Electronic Commerce}}. \bibinfo{pages}{324--330}.
\newblock

\vfill\eject

\bibitem[\protect\citeauthoryear{Jindal and Liu}{Jindal and Liu}{2007}]%
        {jindal2007analyzing}
\bibfield{author}{\bibinfo{person}{Nitin Jindal} {and} \bibinfo{person}{Bing
  Liu}.} \bibinfo{year}{2007}\natexlab{}.
\newblock \showarticletitle{Analyzing and Detecting Review Spam}. In
  \bibinfo{booktitle}{{\em Proceedings of the 7th IEEE International Conference
  on Data Mining}}. \bibinfo{pages}{547--552}.
\newblock

\bibitem[\protect\citeauthoryear{Lee, Hosanagar, and Tan}{Lee
  et~al\mbox{.}}{2015}]%
        {lee2015follow}
\bibfield{author}{\bibinfo{person}{Young-Jin Lee}, \bibinfo{person}{Kartik
  Hosanagar}, {and} \bibinfo{person}{Yong Tan}.}
  \bibinfo{year}{2015}\natexlab{}.
\newblock \showarticletitle{Do I Follow my Friends or the Crowd? Information
  Cascades in Online Movie Ratings}.
\newblock \bibinfo{journal}{{\em Management Science\/}} \bibinfo{volume}{61},
  \bibinfo{number}{9} (\bibinfo{year}{2015}), \bibinfo{pages}{2241--2258}.
\newblock

\bibitem[\protect\citeauthoryear{Leskovec and Krevl}{Leskovec and
  Krevl}{2016}]%
        {leskovec2016snap}
\bibfield{author}{\bibinfo{person}{Jure Leskovec} {and} \bibinfo{person}{Andrej
  Krevl}.} \bibinfo{year}{2016}\natexlab{}.
\newblock \showarticletitle{SNAP Datasets: Stanford Large Network Dataset
  Collection (2014)}.
\newblock \bibinfo{journal}{{\em http://snap.stanford.edu/data\/}}
  (\bibinfo{year}{2016}).
\newblock


\bibitem[\protect\citeauthoryear{McAuley, Leskovec, and Jurafsky}{McAuley
  et~al\mbox{.}}{2012}]%
        {McAuley2012LearningAA}
\bibfield{author}{\bibinfo{person}{Julian~J McAuley}, \bibinfo{person}{Jure
  Leskovec}, {and} \bibinfo{person}{Daniel Jurafsky}.}
  \bibinfo{year}{2012}\natexlab{}.
\newblock \showarticletitle{Learning Attitudes and Attributes from Multi-Aspect
  Reviews}.
\newblock \bibinfo{journal}{{\em Proceedings of the 12th IEEE International
  Conference on Data Mining\/}} (\bibinfo{year}{2012}),
  \bibinfo{pages}{1020--1025}.
\newblock


\bibitem[\protect\citeauthoryear{Muchnik, Aral, and Taylor}{Muchnik
  et~al\mbox{.}}{2013}]%
        {muchnik2013social}
\bibfield{author}{\bibinfo{person}{Lev Muchnik}, \bibinfo{person}{Sinan Aral},
  {and} \bibinfo{person}{Sean~J Taylor}.} \bibinfo{year}{2013}\natexlab{}.
\newblock \showarticletitle{Social Influence Bias: A Randomized Experiment}.
\newblock \bibinfo{journal}{{\em Science\/}} \bibinfo{volume}{341},
  \bibinfo{number}{6146} (\bibinfo{year}{2013}), \bibinfo{pages}{647--651}.
\newblock


\bibitem[\protect\citeauthoryear{Roider and Voskort}{Roider and
  Voskort}{2016}]%
        {roider2016reputational}
\bibfield{author}{\bibinfo{person}{Andreas Roider} {and}
  \bibinfo{person}{Andrea Voskort}.} \bibinfo{year}{2016}\natexlab{}.
\newblock \showarticletitle{Reputational Herding in Financial Markets: A
  Laboratory Experiment}.
\newblock \bibinfo{journal}{{\em Journal of Behavioral Finance\/}}
  \bibinfo{volume}{17}, \bibinfo{number}{3} (\bibinfo{year}{2016}),
  \bibinfo{pages}{244--266}.
\newblock


\bibitem[\protect\citeauthoryear{Rosenbaum}{Rosenbaum}{2010}]%
        {rosenbaum2002observational}
\bibfield{author}{\bibinfo{person}{Paul~R Rosenbaum}.}
  \bibinfo{year}{2010}\natexlab{}.
\newblock \bibinfo{booktitle}{{\em Design of Observational Studies}}.
\newblock \bibinfo{publisher}{Springer}.
\newblock


\bibitem[\protect\citeauthoryear{Scharfstein and Stein}{Scharfstein and
  Stein}{1990}]%
        {scharfstein1990herd}
\bibfield{author}{\bibinfo{person}{David~S Scharfstein} {and}
  \bibinfo{person}{Jeremy~C Stein}.} \bibinfo{year}{1990}\natexlab{}.
\newblock \showarticletitle{Herd Behavior and Investment}.
\newblock \bibinfo{journal}{{\em The American Economic Review\/}}
  \bibinfo{volume}{80}, \bibinfo{number}{3} (\bibinfo{year}{1990}),
  \bibinfo{pages}{465--479}.
\newblock


\bibitem[\protect\citeauthoryear{Wang and Wang}{Wang and Wang}{2014}]%
        {wang2014amazon}
\bibfield{author}{\bibinfo{person}{Ting Wang} {and} \bibinfo{person}{Dashun
  Wang}.} \bibinfo{year}{2014}\natexlab{}.
\newblock \showarticletitle{Why Amazon's Ratings Might Mislead You: The Story
  of Herding Effects}.
\newblock \bibinfo{journal}{{\em Big Data\/}} \bibinfo{volume}{2},
  \bibinfo{number}{4} (\bibinfo{year}{2014}), \bibinfo{pages}{196--204}.
\newblock


\bibitem[\protect\citeauthoryear{Yoav and Shchori-Bachrach}{Yoav and
  Shchori-Bachrach}{1973}]%
        {yoav1973process}
\bibfield{author}{\bibinfo{person}{Kislev Yoav} {and} \bibinfo{person}{Nira
  Shchori-Bachrach}.} \bibinfo{year}{1973}\natexlab{}.
\newblock \showarticletitle{The Process of an Innovation Cycle}.
\newblock \bibinfo{journal}{{\em American Journal of Agricultural Economics\/}}
  \bibinfo{volume}{55}, \bibinfo{number}{1} (\bibinfo{year}{1973}),
  \bibinfo{pages}{28--37}.
\newblock


\end{thebibliography}


\end{document}